# Formal deduction of the Saint-Venant-Exner model including arbitrarily sloping sediment beds and associated energy

E.D. Fernández-Nieto [*], T. Morales de Luna [†]
G. Narbona-Reina[*], J. D. Zabsonré [‡]

June 18, 2015

## Abstract

In this work we present a deduction of the Saint-Venant-Exner model through an asymptotic analysis of the Navier-Stokes equations. A multi-scale analysis is performed in order to take into account that the velocity of the sediment layer is smaller than the one of the fluid layer. This leads us to consider a shallow water type system for the fluid layer and a lubrication Reynolds equation for the sediment one. This deduction provides some improvements with respect to the classical Saint-Venant-Exner model: (i) the deduced model has an associated energy. Moreover, it allows us to explain why classical models do not have an associated energy and how to modify them in order to recover a model with this property. (ii) The model incorporates naturally a necessary modification that must be taken into account in order to be applied to arbitrarily sloping beds. Furthermore, we show that this modification is different of the ones considered classically, and that it coincides with a classical one only if the solution has a constant free surface. (iii) The deduced solid transport discharge naturally depends on the thickness of the moving sediment layer, what allows to ensure sediment mass conservation. Moreover, we include a simplified version of the model for the case of quasi-stationary regimes. Some of these simplified models correspond to the generalization of classical ones such as Meyer-Peter&Müller and Ashida-Michiue models. Three numerical tests are presented to study the evolution of a dune for several definition of the repose angle, to see the influence of the proposed definition of the effective shear stress in comparison with the classical one, and by comparing with experimental data.

---

[*]Dpto. Matemática Aplicada I. ETS Arquitectura - Universidad de Sevilla. Avda. Reina Mercedes N. 2. 41012-Sevilla, Spain. (edofer@us.es), (gnarbona@us.es)

[†]Dpto. de Matemáticas. Universidad de Córdoba. Campus de Rabanales. 14071 Córdoba, Spain (tomas.morales@uco.es)

[‡]Unité de Formation et de Recherche en Sciences et Techniques, Département de Mathématiques, Université Polytechnique de Bobo-Dioulasso, Bobo-Dioulasso, Burkina Faso. (jzabsonre@gmail.com)

# Contents

# 1 Introduction

The Saint-Venant-Exner system (see [17]) is generally used to model the bedload transport in rivers, lakes and coastal areas. Sediment transport is usually divided into three types: surface creep, saltation and suspension. Surface creep is defined as the type of transport where sediment grains roll or slide along the bed. Saltation transport is defined as the type of transport where single grains jump over the bed a length proportional to their diameter, losing for instants the contact with the soil. Sediment is suspended when the flux is intense enough so that the sediment grains reach height over the bed. There is not a clear distinction between surface creep and saltation, so that these types of transport are usually called bedload transport. At low Froude numbers, the bedload is the dominating transport mechanism which is the regime under study in this paper.

The Saint-Venant-Exner system (SVE in what follows) is defined in terms of a hydrodynamical component coupled with a morphodynamical one. The hydrodynamical component in most cases is modeled by Saint-Venant system. The equation that describe the morphodynamical component is the well known Exner equation, that is a continuity equation. Such system can be written under the following form:

$$
\begin{cases}
\partial_t h_1 + \mathrm{div}_x q_1 = 0, \\
\partial_t q_1 + \mathrm{div}_x \left( \dfrac{q_1 \otimes q_1}{h_1} \right) + \nabla_x \left( \dfrac{1}{2} g h_1^2 \right) + g h_1 \nabla_x (h_2 + b) + \tau/\rho_1 = 0, \\
\partial_t h_2 + \mathrm{div}_x q_b = 0,
\end{cases}
\tag{1}
$$

where $x = (x_1, x_2)$ is the horizontal spacial coordinates, $t$ represents the time variable, $q_1 = h_1(x,t) u_1(x,t)$ represents the water discharge, $h_1(x,t)$ the height of the fluid and $u_1 = (u_{1\,[1]}, u_{1\,[2]})$ its horizontal velocity. $\tau$ is the shear stress at the bottom and $\rho_1$ the density of the fluid. The unknown function $h_2 = h_2(x,t)$ is the thickness of the sediment layer (see Figure 1) and $q_b$ denotes the solid transport discharge. $g$ is the gravity constant and $b$ is the fixed bottom, usually called the bedrock layer. In what follows we shall denote $\eta = b + h_2$, being $z = \eta(x,t)$ the sediment bed surface.

To close the system, it is necessary to define the solid transport discharge $q_b$. Several formulae for $q_b$ can be found in the literature. For example the classical formula proposed by Grass [29] assumes that the movement of the sediment begins at the same time as for the fluid and both move in the same direction. It is defined by $q_b = A_g |u_1|^{m_g - 1} u_1$ where $A_g$ is a constant which takes into account the grain size and the kinematic viscosity and $m_g$ is a positive real number, such that $1 \le m_g \le 4$. Nevertheless, for practical applications, some other type of formulae has been proposed in the literature, for instance, by Meyer-Peter & Müller [39], Van Rijn's [53], Einstein [16], Nielsen [43], Fernández-Luque & Van Beek [6, 20], Ashida & Michiue [1], Engelund & Fredsoe [18], Kalinske [32] or Charru [10]. Such formulae are usually presented in nondimensional form and can be written as follows,

$$
\frac{q_b}{Q} = \mathrm{sgn}(\tau) \frac{k_1}{(1 - \varphi)} \theta^{m_1} (\theta - k_2 \theta_c)_+^{m_2} \left( \sqrt{\theta} - k_3 \sqrt{\theta_c} \right)_+^{m_3},
\tag{2}
$$

where $Q$ represents the characteristic discharge, $Q = d_s \sqrt{g(1/r - 1) d_s}$, $r$ is the density ratio, $r = \rho_1/\rho_2$, being $\rho_2$ the density of the sediment particles and $d_s$ the mean diameter of the sediment particles. $\varphi$ is the averaged porosity.

In classical SVE models the sign of $q_b$ coincides with the sign of $\tau$, the shear stress at the bottom. It is usually defined as $\tau = \rho_1 g h_1 S_f$, being $S_f$ the friction term. $S_f$ can be set by different empirical laws such as the Darcy-Weisbach ($S_f = f u_1 |u_1|/8 g h_1$, where $f$ is the Darcy-Weisbach coefficient) or Manning formulae ($S_f = n^2 u_1 |u_1|/h_1^{4/3}$, where $n$ is the Manning coefficient). In general we can write $\tau = \rho_1 g \zeta(h_1) u_1 |u_1|$, being $\zeta(h_1)$ a function depending on the considered friction law.

The Shields stress, $\theta$, represents the ratio between the agitating and the stabilizing forces on a sediment grain in the bed,

$$\theta = \frac{|\tau| d_s^2}{g(\rho_2 - \rho_1) d_s^3},\tag{3}$$

and $\theta_c$ is the critical Shields stress for incipient motion.

The constants $k_l$, $m_l$, $l = 1, 2, 3$ are positive real numbers depending on the model. Being usually at least one of the parameters $m_1$, $m_2$ or $m_3$ equals to zero. For example, Meyer-Peter & Müller's model is defined by

$$\frac{q_b}{Q} = \text{sgn}(\tau) \frac{8}{(1 - \varphi)} (\theta - \theta_c)_+^{3/2},\tag{4}$$

and Ashida & Michiue's model is defined by

$$\frac{q_b}{Q} = \text{sgn}(\tau) \frac{17}{(1 - \varphi)} (\theta - \theta_c)_+ (\sqrt{\theta} - \sqrt{\theta_c}).\tag{5}$$

Finally, by $(\,\cdot\,)_+$ we denote the positive part. In equation (2) the positive part implies that the sediment moves when the modulus of the shear stress is bigger than a given critical value.

Although the classical SVE model is largely used, it presents several disadvantages:

(i) The SVE model has not a dissipative energy equation associated to the system.

(ii) Solid transport discharge formulae are derived by using the hypothesis of nearly horizontal sediment beds, that is, $\nabla_x \eta \approx 0$.

(iii) Solid transport flux is independent of the thickness of the sediment layer. Thus, the mass conservation property for the sediment given by third equation in (1) may fail (see [40]).

Concerning the first item, we should remark that there exist in the literature some simpler solid transport formulae for which the corresponding SVE model has an associated dissipative energy equation (see for example [36]). Nevertheless, up to our knowledge, no general result exists in the bibliography in this sense.

The second item implies that classical formulae cannot be used in several problems of interest (see [33]) because they are derived by using the hypothesis of nearly horizontal sediment beds.

As mentioned before, the Shields parameter is the coefficient between agitating forces and the stabilizing forces. Classical formulae consider that the only agitating force is the bottom shear stress, concretely $|\tau| d_s^2$. Nevertheless, in the experiments presented by Lysne in [37] it can be seen that gravity is another contributing factor as an agitating force (see also [27]) for sloped sediment beds. Then, it is necessary to take into account gravitational forces in order to obtain a solid transport discharge that can be applied in arbitrarily sloping beds.

This has been done in the literature in several ways. For instance, the simplest way to take into account the sediment bed slope in the definition of the solid transport discharge is to include a diffusion term. Engelund and Hansen proposed in [19] a formula that can be written under the following form,

$$q_b = k|u_1|^m \left( \frac{u_1}{|u_1|} - c \, \nabla_x \eta \right),$$
(6)

$k$, $m$, and $c$ being constant parameters of the model (see also [58] and [59]). Equation (6) can be seen as a modification of the Grass model. An adaptation of this formula for curved channels was proposed by Struiksma et al. in [57]:

$$q_b = k|u_1|^m \left( 1 - c \, \partial_s (b + h_2) \right) \operatorname{sgn} \left( u_1 - \frac{1}{f_s \, \theta} \nabla_x \eta \right),$$

$s$ being the streamwise coordinate and $f_s$ the shape factor of the grains (see [57] for more details). Note that this definition implies that the direction of the sediment transport does not coincides with the direction of the velocity of the fluid. The direction is determinated by the sign of the vector $(u_1 - \frac{1}{f_s \theta} \nabla_x \eta)$. Note that

$$\operatorname{sgn} \left( u_1 - \frac{1}{f_s \, \theta} \nabla_x \eta \right) = \frac{u_1 - \frac{1}{f_s \, \theta} \nabla_x \eta}{\left| u_1 - \frac{1}{f_s \, \theta} \nabla_x \eta \right|} = \left( \begin{array}{c} \cos \alpha \\ \sin \alpha \end{array} \right),$$

$\alpha$ being the angle of the transport direction, where

$$\tan \alpha = \frac{u_{1\,[2]} - \frac{1}{f_s \, \theta} \partial_{x_2} \eta}{u_{1\,[1]} - \frac{1}{f_s \, \theta} \partial_{x_1} \eta}.$$
(7)

Let us remark that in [57] the direction of the sediment transport is defined in terms of its angle, given by (7), rather than defining it in terms of the sign vector. Authors also include a correction of the angle due to the transversal velocity in curves. For the sake of brevity we do not include a discussion on the modification of the transport angle in curved channels, since it is not the aim of this paper.

A more used extension of classical formulae to arbitrarily sloping sediment beds is to consider a modification of the critical Shields parameter, replacing $\theta_c$ by $\widehat{\theta}_c$ (see [27] and [20]). For the scalar case, 1D flows, this modification can be written as follows:

$$\widehat{\theta}_c = \theta_c \left( 1 + \frac{\operatorname{sgn}(\tau)}{\tan \delta} \partial_x \eta \right) = \theta_c + \vartheta \operatorname{sgn}(\tau) \, \partial_x \eta, \quad \text{where} \quad \vartheta = \frac{\theta_c}{\tan \delta}.$$
(8)

In the works of Kovacs and Parker [33], Seminara et al. [56] and Parker et al. [47] several extensions for the vectorial case are presented, where the computation of $\widehat{\theta}_c$ takes into account lateral slopes.

Note that usually in the definition of the modified critical Shields parameter we do not find the sign of $\tau$. This is due to the fact that formulae are usually presented for the case of positive velocities only.

In previous definition $\tan\delta$ is the friction coefficient corresponding to the internal friction angle of the material. Typical values of $\vartheta$ and $\theta_c$ are, $\vartheta = 0.1$ and $\theta_c = 0.047$, what implies that $\theta_c/\vartheta = \tan\delta$, with $\delta \approx 25^{\circ}$, as proposed by Fredsœ in [27]. The angle $\delta = 25^{\circ}$ is lower than the repose angle close to $32^{\circ}$, although lower values have also been suggested (see [10] and references therein).

Let us remark that the definition of $\widehat{\theta_c}$ in (8) is based in two arguments: first, gravitational forces due to the sediment bed slope are incorporated as agitating forces in the definition of the effective Shields parameter,

$$\theta_{\text{eff}} = \frac{|\tau_{\text{eff}}|d_s^2}{g(\rho_2 - \rho_1)d_s^3} \quad \text{with} \quad \tau_{\text{eff}} = \tau - \vartheta(\rho_2 - \rho_1)gd_s\partial_x(b + h_2), \tag{9}$$

where $\tau_{\text{eff}}$ is the effective shear stress. Second, rather than replacing $\theta$ by $\theta_{\text{eff}}$, it is usually assumed that this is equivalent to replacing $\theta_c$ by $\widehat{\theta_c}$, defined by (8) (see [27]). More explicitly, it is usually assumed that $(\theta_{\text{eff}} - \theta_c) = (\theta - \widehat{\theta_c})$.

Nevertheless, this second assumption is not true in general,

$$\theta_{\text{eff}} - \theta_c \neq \frac{|\tau|d_s^2 - \vartheta\,\text{sgn}(\tau)g(\rho_2 - \rho_1)d_s^3\partial_x(b + h_2)}{g(\rho_2 - \rho_1)d_s^3} - \theta_c = \theta - \widehat{\theta_c}.$$

The problem arises from the fact that the absolute value (or the norm in the vectorial case) is neglected in the definition of $\theta_{\text{eff}}$ (see for example [10] and [27]), which should be taken into account, depending on the sign of $\tau$ and $\tau_{\text{eff}}$.

In fact, Fowler et al. proposed in [26] a modification of the Meyer-Peter & Müller formula that consist in replacing $\theta$ by $\theta_{\text{eff}}$, instead of replacing $\theta_c$ by $\widehat{\theta_c}$. The model proposed by Fowler et al. can be written as follows:

$$\frac{q_b}{Q} = 8\,\text{sgn}(\tau_{\text{eff}})\frac{h_2}{\bar{h}_2}\left(\theta_{\text{eff}} - \theta_c\right)_+^{3/2},$$

where $\bar{h}_2$ is an averaged value of the thickness of the sediment layer. The fact of introducing an explicit dependence on $h_2$ in the formula is also interesting. Indeed, this allows to ensure the mass conservation property for the sediment layer (see also [40]), which was the second issue noted previously as a disadvantage for classical SVE models.

Related to this problem, in [23], Fernández-Nieto et al. introduce a modified general definition of the solid transport discharge for SVE models that takes into account the thickness of the sediment layer. Then mass conservation is ensured. Moreover, the proposed formula has the advantage that it reduces to a classical solid transport discharge formula in the case of quasi-uniform regimes. This is in fact the regime where usually classical formulae are derived.

The main objective of this paper is to show that the SVE model can be deduced through an asymptotic analysis of the Navier-Stokes equations. Moreover, we obtain that following this process some improvements on the classical SVE model are reached: (i) We obtain that the deduced model verifies exactly a dissipative energy equation. This deduction and the proof of energy allows us to understand why classical models do not have an associated energy. Moreover, it allows to prove that the classical SVE model can have an associated energy by introducing a simple modification. (ii) In the deduction process of the model the pressure terms introduce a modification that must be taken into account if we consider applications where the sediment bed is not nearly horizontal. We also see that this modification coincides with some of the alternatives proposed in the literature for some special cases. (iii) The solid transport flux depends on the thickness of the moving sediment layer, then mass conservation is ensured.

The paper is organized as follows: In Section 2 we present the 3D system of equations considered as starting point from which the models are deduced, the proposed models and the results on the associated energy. The formal deduction of the models by an asymptotic analysis from the Navier-Stokes equations is detailed in Appendix A. This is done by developing a multi-scale analysis in space and time. Section 3 is devoted to numerical tests. In the first test we study the evolution of a dune for different values of the the repose angle. The purpose of the second test is to compare the influence of the deduced modification in the definition of the effective shear stress respect to classical ones. A comparison with experimental data is presented in the third test. Finally, the conclusions are presented in Section 4.

## 2   The asymptotic Saint-Venant-Exner models

Following an asymptotic analysis, we derive two main models which are presented here (see Appendix A for details). The difference between these models is the friction law used at the fluid and sediment interaction level. In Subsection 2.1 we summarize the starting 3D system of equations and the hypothesis considered for the derivation of the models. The models deduced in this work are shown in Subsection 2.2. First, they are presented with the same notation as they are deduced (see Appendix A). Secondly, they are rewritten in terms of the Shields parameter or the effective Shields parameter. The associated energy for these models is presented in Subsection 2.3. Moreover, we include in Subsection 2.4 a result that justifies that classical SVE model may have an associated dissipative energy provided a second order correction is made in the friction term appearing in the momentum equation.

### 2.1   The 3D initial system

We consider two immiscible layers of different materials with different physical properties: velocity, pressure, density and viscosity. The two layers are related through the interaction terms at the internal interface levels. In the following subsections the starting systems of equations, the definition of the physical domain and the considered hypothesis for the deduction of the models proposed in this paper are specified.

### 2.1.1 Physical domain and governing equations

We consider a cartesian coordinate system where $x$ represents the horizontal 2D direction and $z$ the vertical one. Let us define the physical domain for the fluid and sediment layers by $\Omega_1(t)$ and $\Omega_2(t)$ respectively; $t$ being the time variable. Usually in the context of bedload transport it is assumed that the sediment domain can be decomposed into two layers: one that moves due to the action of the convection of the upper fluid, the mobile sediment layer with thickness $h_m$, and a second one composed of sediment that is not moving but is susceptible to come into motion, with thickness $h_f$. This leads us to define four boundaries in the domain (see Figure 1):

$$\Gamma_s = \{(x, z) \in \mathbb{R}^3 / x \in \omega, \ z = \eta_s(x, t)\};$$
$$\Gamma_{1,2} = \{(x, z) \in \mathbb{R}^3 / x \in \omega, \ z = \eta(x, t)\};$$
$$\Gamma_f = \{(x, z) \in \mathbb{R}^3 / x \in \omega, \ z = \eta_f(x, t)\};$$
$$\Gamma_b = \{(x, z) \in \mathbb{R}^3 / x \in \omega, \ z = b(x)\}.$$

where $\omega$ is a domain in $\mathbb{R}^2$. The water free surface is defined by $z = \eta_s(x, t)$, where $\eta_s(x, t) = h_1(x, t) + h_2(x, t) + b(x)$ and the fluid/sediment interface by $z = \eta(x, t)$, where $\eta(x, t) = b(x) + h_2(x, t)$. $h_1(x, t)$ denotes the height of the water column. The sediment layer is decomposed as $h_2(x, t) = h_m(x, t) + h_f(x, t)$, and then the internal sediment interface is $z = \eta_f(x, t)$, where $\eta_f(x, t) = b(x) + h_f(x, t)$ (see Figure 1).

Thus, we consider a time-dependant domain $\Omega(t) = \Omega_1(t) \cup \Omega_2(t) \cup \Gamma_b \cup \Gamma_{1,2}(t) \cup \Gamma_s(t)$, being:

$$\Omega_1(t) = \{(x, z) \in \mathbb{R}^3 / x \in \omega, \ \eta(x, t) < z < \eta_s(x, t)\};$$
$$\Omega_2(t) = \Omega_{2,f}(t) \cup \Omega_{2,m}(t);$$
$$\text{where}$$
$$\Omega_{2,f}(t) = \{(x, z) \in \mathbb{R}^3 / x \in \omega, \ b(x) < z < \eta_f(x, t)\};$$
$$\Omega_{2,m}(t) = \{(x, z) \in \mathbb{R}^3 / x \in \omega, \ \eta_f(x, t) < z < \eta(x, t)\};$$

Moreover, let us denote by $T_m$ the mass transference between the static and the mobile sediment domains, $\Omega_{2,f}$ and $\Omega_{2,m}$. $T_m$ is defined as the difference between the erosion rate ($\dot{z}_e$) and the deposition rate ($\dot{z}_d$), $T_m = \dot{z}_e - \dot{z}_d$ (see [23] and Section 2.2.3).

As a general rule in notation, we will use the subscript 1 to denote the upper layer (fluid) and the subscript 2 for the lower layer (sediment). We denote by

$$v_i = (u_i, w_i)$$

the velocity field for each layer with $u_i = (u_{i[1]}, u_{i[2]})$ the 2D horizontal velocity. We denote by $\rho_i$ the density and $p_i$ the pressure. Moreover, $\mu_i$ and $\nu_i = \mu_i/\rho_i$, denote the dynamic and kinematic viscosity coefficients respectively, for $i = 1, 2$.

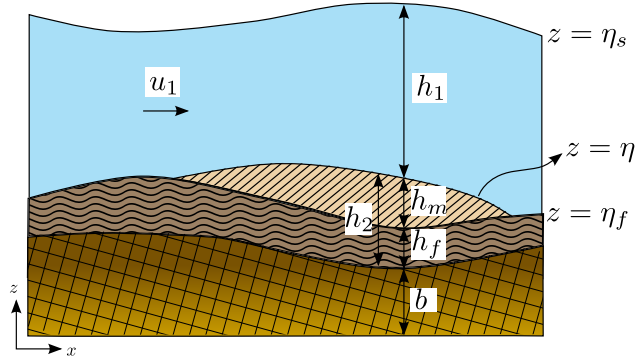

Figure 1: Sketch of the domain for the fluid-sediment problem

For each layer $(i = 1, 2)$, we start from the 3D Navier-Stokes equations for incompressible fluid and sediment components:

$$\begin{cases} \rho_i \partial_t v_i + (\rho_i v_i \nabla) v_i - \text{div}(\sigma_i) = -\rho_i \vec{g} \\ \text{div}(v_i) = 0 \end{cases} \tag{10}$$

In this system $\vec{g}$ represents the gravitational vector and $\sigma_i$ the stress tensor associated to each layer.

To complete this system, we must give the stress tensor expressions, the interactions at the internal interface levels ($\Gamma_{1,2}$ and $\Gamma_f$), as soon as boundary and kinematic conditions. They are specified in the following subsection.

### 2.1.2 Closures

**Stress tensors**

We shall define the stress tensors as follows:

$$\sigma_i = \sigma_i' - p_i \text{Id}, \quad \text{for } i = 1, 2;$$

where $\sigma_i'$ is the deviatoric part of the stress tensor.

We assume that the stress tensor for the fluid layer follows a Newtonian form with constant dynamic viscosity $\mu_1$ and then $\sigma_1'$ is given by:

$$\sigma_1' = 2\mu_1 D(v_1),$$

$D(v)$ being the rate of deformation tensor, $D(v) = \frac{1}{2}(\nabla v + \nabla^t v)$.

For the sediment layer we consider a non-Newtonian rheology. Recent works have been devoted to demonstrate through experimental results the resemblance between the bed load transport phenomena and the granular flows behavior [45, 46, 2, 52]. Ouriemi et al. proposed in [45] a two-phase model for bed load transport in laminar flows. Where a Newtonian law for the fluid phase and a frictional rheology for the particulate phase, namely a Coulomb type friction is considered. Chauchat et al. (see [7], [8]) study a three-dimensional two-phase model where a Drucker-Prager rheology describe the stress tensor associated to the particulate phase. This definition is consistent with the Coulomb friction model proposed in [45]. A comparison with experimental data for bed load transport can be seen in [2].

Following these references we define

$$\sigma_2' = 2\mu_2 D(v_2),$$

with a non-constant viscosity $\mu_2$, given by a Drucker-Prager model. By using a regularization of the Drucker-Prager model (see [35]), we can assume that $\mu_2$ is expressed as a function of $D(v_2)$ and $p_2$.

Nevertheless, let us remark that in order to deduce a first order Reynolds-type model for the sediment layer, we do not need further specifications about this rheology. In fact, the depth-averaged model will be written in terms of the boundary conditions, that is, the Coulomb friction law, which is compatible with the Drucker-Prager rheology.

**Friction laws**

We must define the friction laws at the mobile-static sediment interface and at the fluid-sediment interface.

∘ *Friction law at the mobile-static sediment interface*

As explained before, a Coulomb friction condition at the interface between the static and the moving sediment particles, $z = \eta_f(x, t)$, is considered. Denoting by $N_f$ the unitary normal vector to the interface $\Gamma_f$ and by $\delta$ the repose angle, we write the Coulomb friction law as

$$(\sigma_2 N_f)_\mathcal{T} = -\Big(\text{sgn}(u_2) \tan \delta \left((\sigma_1 - \sigma_2) N_f\right) \cdot N_f\Big)_{|z = \eta_f}, \tag{11}$$

where $N_f = (-\nabla_x \eta_f, \ 1)^t / \sqrt{1 + |\nabla_x \eta_f|^2}$ and the subscript $(\cdot)_\mathcal{T}$ denotes the tangent component of a vector.

Let us remark that the use of a Coulomb friction can also be interpreted as a mechanism to approximate the collision effects of saltating grains in the computation of the bedload transport formula (see [33]).

∘ *Friction law at the fluid-sediment interface*

At the fluid-sediment interface the interaction between fluid and sediment is defined through a friction force:

$$(\sigma_1 N_\eta)_\mathcal{T} = (\sigma_2 N_\eta)_\mathcal{T} = (\text{fric}, 0)^t, \tag{12}$$

where $N_\eta = (-\nabla_x \eta, \ 1)^t / \sqrt{1 + |\nabla_x \eta|^2}$ is the normal vector at the interface pointing from layer 2 to layer 1.

In particular, we consider two classical friction laws for which we obtain two different models that can be written under the same structure.

- Linear friction law:

$$\text{fric} = C \, (u_1 - u_2)_{|z=\eta},\tag{13}$$

  where the coefficient $C$ has velocity dimension. Moreover, taking into account the results presented in [42] (see also Remark A.2), $C$ can be assumed proportional to $h_m$. Following the analysis of Seminara et al. [56], the drag coefficient is proportional to $\tan(\delta)/\theta_c$. That is, inversely proportional to $\vartheta$, defined in (8). Taking into account these remarks we may define

$$C = \frac{(1/r - 1) g h_m}{\vartheta \sqrt{(1/r - 1) g d_s}}.\tag{14}$$

- Quadratic friction law:

$$\text{fric} = C_1 \, |(u_1 - u_2)_{|z=\eta}| \, (u_1 - u_2)_{|z=\eta}.\tag{15}$$

  In this case $C_1$ must be adimensional, so taking into account previous arguments on the definition of the drag coefficient, we can define

$$C_1 = \frac{h_m}{\vartheta \, d_s}.\tag{16}$$

**Boundary and kinematic conditions**

The following boundary and kinematic conditions are imposed at each interface to complete the system:

- At the free surface, $z = \eta_s(t, x) = b(x) + h_2(x, t) + h_1(x, t)$:

  - The surface tension condition: $(\sigma_1 \cdot N_s)_n = 0$ where $N_s = \frac{1}{\sqrt{1 + |\nabla_x \eta_s|^2}} (-\nabla_x \eta_s, \ 1)^t$ is the unitary outward normal vector to the free surface and the subscript $n$ denotes the normal component.

  - The kinematic condition: $\partial_t \eta_s = v_1 \cdot N_s$.

- At the fluid/sediment interface, $z = \eta(t, x) = b(x) + h_2(x, t)$:

  - The kinematic conditions corresponding to both velocities:

  $$\partial_t \eta = v_1 \cdot N_\eta = v_2 \cdot N_\eta.$$

  - The continuity of the normal component of the tensors: $(\sigma_1 \cdot N_\eta)_n = (\sigma_2 \cdot N_\eta)_n$.

– The friction law defined by (12)-(16).

• At the internal sediment interface, $z = \eta_f(t, x) = b(x) + h_f(x, t)$:

– The conservation of the sediment mass,

$$\partial_t \eta_f = v_2 \cdot N_f - T_m;$$

$T_m$ being the mass transference term (see Section 2.2.3).

– The Coulomb friction law defined by (11).

• At the bottom, $z = b(x)$:

– The no penetration condition: $v_2 \cdot N_b = 0$, where the unitary normal vector to the bottom is $N_b = (-\nabla_x b, \ 1)^t / \sqrt{1 + |\nabla_x b|^2}$.

## 2.2   Proposed models

In this section we present the final models obtained through an asymptotic analysis of the 3D system (10). Following the work performed in [24], we derive a mathematical two dimensional SVE type model for bed load transport. Thus, the models are composed by three equations: the first two equations represent the Saint-Venant system used for modeling the upper fluid layer and the third one describes the evolution of the moving bed through a lubrication Reynolds equation.

As it is classically considered in such kind of models, we take into account two different time scales for the hydrodynamics and the sediment evolution. Therefore, the ratio between the hydrodynamic and morphodynamic time scales, defined by $\varepsilon^2$, being small. We obtain first order models, so we neglect the terms of order $\varepsilon^2$ as commonly done to obtain SVE type models, (see [28] and references therein).

The two models obtained depends on the considered friction law at the fluid-sediment interface, given by (13) and (15) respectively. The complete derivation of these models is detailed in Appendix A.

First, we present the models deduced under a unified formulation. Then, we analyze each of them in order to relate the sediment velocity with the classical motion threshold in terms of the Shields parameter and the effective Shields parameter. Finally, the closure of the systems is discussed and simplified models corresponding to a quasi stationary regime will be presented. From now on, we use the superscript $(LF)$ to denote the properties of the model deduced with a linear friction law, and $(QF)$ for the model using a quadratic friction law.

The model deduced in this paper can be written under the form of a SVE system form

as follows:

$$\begin{cases} \partial_t h_1 + \mathrm{div}_x q_1 = 0, \\[2mm] \partial_t q_1 + \mathrm{div}_x(h_1(u_1 \otimes u_1)) + \dfrac{1}{2}g\nabla_x h_1^2 + gh_1\nabla_x(b+h_2) + \dfrac{gh_m}{r}\mathcal{P} = 0, \\[2mm] \partial_t h_2 + \mathrm{div}_x\left(h_m\, v_b\, \sqrt{(1/r-1)gd_s}\right) = 0, \\[2mm] \partial_t h_f = -T_m. \end{cases} \tag{17}$$

with

$$\mathcal{P} = \nabla_x(rh_1 + h_2 + b) + (1-r)\mathrm{sgn}(u_2)\tan\delta. \tag{18}$$

and the velocity $v_b$ defined for each friction law as

$$v_b^{(LF)} = \frac{1}{\sqrt{(1/r-1)gd_s}}u_1 - \frac{\vartheta}{1-r}\mathcal{P}, \tag{19}$$

$$v_b^{(QF)} = \frac{1}{\sqrt{(1/r-1)gd_s}}u_1 - \left(\frac{\vartheta}{1-r}\right)^{1/2}|\mathcal{P}|^{1/2}\mathrm{sgn}(\mathcal{P}). \tag{20}$$

We can write these velocities in terms of the Shields parameter. For this aim, we use now the expression of the shear stress and the definitions of the friction coefficients $C$ and $C_1$ given by (14) and (16). We obtain,

$$v_b^{(LF)} = \mathrm{sgn}(\tau_{\mathrm{eff}}^{(LF)})(\theta_{\mathrm{eff}}^{(LF)} - \theta_c)_+,$$

where $\tau_{\mathrm{eff}}^{(LF)}$ and $\theta_{\mathrm{eff}}^{(LF)}$ are defined by (24) and (25), respectively, for the case of a linear friction law. For the case of a quadratic friction law $v_b^{(QF)}$ is defined by (35). If we consider a linearization of the quadratic friction law (see Subsection 2.2.2), we obtain

$$v_b^{(QF)} = \mathrm{sgn}(\tau_{\mathrm{eff}}^{(QF)})((\theta_{\mathrm{eff}}^{(QF)})^{1/2} - \theta_c^{1/2})_+,$$

where $\tau_{\mathrm{eff}}^{(QF)}$ and $\theta_{\mathrm{eff}}^{(QF)}$ are defined by (32)-(33).

In the following subsections we describe the deduction of previous definitions of $v_b$ for each friction law.

### 2.2.1 Model deduced with the linear friction law (13)-(14)

From the asymptotic approximation at first order, we introduce the shear stress and the Shields parameter as

$$\frac{\tau^{(LF)}}{\rho_1} = Cu_1 \quad \text{and} \quad \theta^{(LF)} = \frac{|\tau^{(LF)}|/\rho_1}{(1/r-1)gd_s} = \frac{C|u_1|}{(1/r-1)gd_s}.$$

Using the definition of the friction coefficient $C$ in equation (14), we have

$$\frac{u_1}{\sqrt{(1/r-1)gd_s}} = \vartheta\,\mathrm{sgn}(u_1)\frac{d_s}{h_m}\,\theta^{(LF)}.$$

Thus, taking into account (18) and the definition of $\vartheta$ in (8), the velocity of the sediment layer reads in this case

$$v_b^{(LF)} = \vartheta\,\mathrm{sgn}(u_1)\frac{d_s}{h_m}\,\theta^{(LF)} - \frac{\vartheta}{1-r}\nabla_x(rh_1+\eta) - \mathrm{sgn}(u_2)\theta_c. \tag{21}$$

∘ *Influence of the Coulomb friction law.*

Note that the sign of the velocity of the sediment layer, $\mathrm{sgn}(u_2)$, has still to be defined. Observe that this coefficient comes from the contribution of the Coulomb friction law at the interface between moving and static sediment particles (see (18)). In order to specify the sign of $u_2$, we remark that Coulomb friction force has the same sign of the net force acting on the sediment. That is,

$$\mathrm{sgn}\left(\frac{\vartheta\,h_m}{1-r}(1-r)\mathrm{sgn}(u_2)\tan\delta\right) = \mathrm{sgn}\left(\frac{h_m u_1}{\sqrt{(1/r-1)gd_s}} - \frac{\vartheta\,h_m}{1-r}\nabla_x(rh_1+h_2+b)\right).$$

Then, using that $\vartheta = \frac{\theta_c}{\tan\delta}$,

$$\mathrm{sgn}(u_2) = \mathrm{sgn}\left(\frac{u_1}{\sqrt{(1/r-1)gd_s}} - \frac{\vartheta}{1-r}\nabla_x(rh_1+h_2+b)\right). \tag{22}$$

Furthermore, Coulomb friction force implies also a threshold in order to stop the motion of the sediment layer when the net forces acting on the sediment is not large enough to compensate the friction force. In this case, we obtain that the velocity of the sediment layer $v_b$ must be zero under the following condition:

$$\left|\frac{h_m u_1}{\sqrt{(1/r-1)gd_s}} - \frac{\vartheta\,h_m}{1-r}\nabla_x(rh_1+h_2+b)\right| \le \frac{\vartheta\,h_m}{1-r}(1-r)\tan\delta = h_m\theta_c. \tag{23}$$

∘ *Link with the effective and the critical Shields parameter.*

The stop criteria (23) gives us the way to find the relationship with the classical motion threshold. Thus, we introduce the modified shear stress including the gravitational forces, called the "effective shear stress", defined in our case as:

$$\frac{\tau_{\mathrm{eff}}^{(LF)}}{\rho_1} = \frac{\kappa\,d_s}{h_m}\frac{\tau^{(LF)}}{\rho_1} - \frac{\vartheta\,gd_s}{r}\nabla_x(rh_1+h_2+b). \tag{24}$$

The corresponding definition of the effective Shields parameter is

$$\theta_{\text{eff}}^{(LF)} = \frac{|\tau_{\text{eff}}^{(LF)}|/\rho_1}{(1/r - 1)gd_s} = \frac{\text{sgn}(\tau_{\text{eff}}^{(LF)})}{(1/r - 1)gd_s} \left( \vartheta \, \text{sgn}(u_1) \frac{d_s}{h_m} \, \theta^{(LF)} - \frac{\vartheta}{1 - r} \nabla_x (rh_1 + h_2 + b) \right). \tag{25}$$

In particular, with these definitions we obtain that $\text{sgn}(\tau_{\text{eff}}^{(LF)}) = \text{sgn}(u_2)$ thanks to (22) and the definition of $\tau^{(LF)}$. As a consequence, from (21), the velocity of the sediment layer can be written as

$$v_b^{(LF)} = \text{sgn}(\tau_{\text{eff}}^{(LF)})(\theta_{\text{eff}}^{(LF)} - \theta_c)_+. \tag{26}$$

Observe that condition (23) coming from the Coulomb friction law is equivalent to the classical one, $\theta_{\text{eff}}^{(LF)} > \theta_c$.

**Remark 2.1 (Comparison with the classical effective shear stress)** *The definition of the effective shear stress (24) can be seen as a generalization of the classical one, given in equation (9) that can be written in vectorial form as follows (see [26]),*

$$\frac{\tau_{eff}}{\rho_1} = \frac{\tau}{\rho_1} - \vartheta(1/r - 1)gd_s \nabla_x (h_2 + b). \tag{27}$$

*There are two main differences: the first one is the factor $(\kappa d_s/h_m)$ multiplying the shear stress, which comes from the fact that the model is expressed in terms of the thickness of the moving sediment layer. The second main difference appears in the second term, that is, we consider $(gd_s \vartheta \nabla_x (rh_1 + h_2 + b)/r)$ instead of $(gd_s \vartheta(1/r - 1)\nabla_x (h_2 + b))$. Nevertheless, both terms are related:*

$$\begin{aligned}
gd_s\vartheta(1/r - 1)\nabla_x (h_2 + b) &= \frac{g \, d_s \, \vartheta}{r} \left( \nabla_x (rh_1 + h_2 + b) - r\nabla_x (h_1 + h_2 + b) \right) \\
&= \frac{g \, d_s \, \vartheta}{r} \nabla (rh_1 + h_2 + b) - g \, d_s \, \kappa \nabla_x (h_1 + h_2 + b).
\end{aligned} \tag{28}$$

*Then, both definitions coincide in the case of constant free surface, $\nabla_x (h_1 + h_2 + b) = 0$, but could be relevant otherwise.*

### 2.2.2 Model deduced with the quadratic friction law (15)-(16)

For the quadratic friction law we have the following definition of the fluid shear stress for the first order approximation

$$\frac{\tau^{(QF)}}{\rho_1} = C_1|u_1|u_1, \quad \text{and then} \quad \theta^{(QF)} = \frac{C_1|u_1|^2}{(1/r - 1)gd_s}, \tag{29}$$

Then, using the definition of $C_1$ in (16) and again (18) and (8), the velocity of the sediment is written as

$$\begin{aligned}
v_b^{(QF)} = \text{sgn}(u_1) \Big( \vartheta \tfrac{d_s}{h_m} \theta^{(QF)} \Big)^{1/2} \\
- \left| \frac{\vartheta}{1 - r} \nabla_x (rh_1 + h_2 + b) + \text{sgn}(u_2)\theta_c \right|^{1/2} \text{sgn} \left( \frac{\vartheta}{1 - r} \nabla_x (rh_1 + h_2 + b) + \text{sgn}(u_2)\theta_c \right).
\end{aligned} \tag{30}$$

Following analogous arguments as in the case of linear friction law, we can define

$$\text{sgn}(u_2) = \text{sgn}(\Phi) \qquad (31)$$

where

$$\Phi = \text{sgn}(u_1)\left(\vartheta\frac{d_s}{h_m}\theta^{(QF)}\right)^{1/2} - \left|\frac{\vartheta}{1-r}\nabla_x(rh_1+h_2+b)\right|^{1/2}\text{sgn}\left(\frac{\vartheta}{1-r}\nabla_x(rh_1+h_2+b)\right), \qquad (32)$$

and similarly to (23), the stop criteria is

$$|h_m\Phi| \le h_m\theta_c.$$

For this model we introduce the effective shear stress as

$$\tau_{\text{eff}}^{(QF)} = g(1/r-1)d_s|\Phi|\Phi \quad \text{and} \quad \theta_{\text{eff}}^{(QF)} = |\Phi|^2, \qquad (33)$$

where $\Phi$ is defined by (32). The former stop criteria, derived directly from the model with the Coulomb friction law, becomes

$$\theta_{\text{eff}}^{(QF)} > \theta_c. \qquad (34)$$

Observe that in this case we are not able to obtain an explicit expression of the velocity $v_b^{(QF)}$ in terms of $\theta_{\text{eff}}^{(QF)}$, due to the presence of the modulus and the sign vectors. Nevertheless, the definition of $v_b^{(QF)}$ must take into account previous criteria. Thus, thanks to (30) and (34), we give

$$v_b^{(QF)} = \begin{cases} v_b^{(QF)} \text{ defined by (30)}, & \text{if } \theta_{\text{eff}}^{(QF)} > \theta_c, \\ 0, & \text{otherwise.} \end{cases} \qquad (35)$$

We may also consider the following approximation, which can be seen as a linearization, of the quadratic friction law (15):

$$\text{fric} = C_1\,|(u_1-u_2^*)_{|z=\eta}|(u_1-u_2^*)_{|z=\eta} + C_1\,|(u_2^*-u_2)_{|z=\eta}|(u_2^*-u_2)_{|z=\eta},$$

being $u_2^*$ the velocity of the system without considering the Coulomb friction law at the internal interface of the sediment layer. In this case, we obtain the following bedload velocity:

$$v_b^{(QF)} = \text{sgn}(\tau_{\text{eff}}^{(QF)})((\theta_{\text{eff}}^{(QF)})^{1/2} - \theta_c^{1/2})_+. \qquad (36)$$

This corresponds to defining a solid transport discharge with a structure similar to that of classical formulae (see for example the Ashida & Michiue's model (5)).

**Remark 2.2** *Definition (33) of the effective shear stress is different of the classical ones. Nevertheless, as in previous case (see equation (24)), the main difference is to consider the term $\nabla_x(rh_1+h_2+b)/(1-r)$ instead of $\nabla_x(h_2+b)$. Note also that a linearization of (33) leads to previous definition of the effective shear stress (24).*

### 2.2.3 Sediment mass transference and simplified models

To close the models, it is necessary to define $T_m$, the mass transference between the moving and the static sediment layers. It is defined in terms of the difference between the erosion rate, $\dot{z}_e$, and the deposition rate $\dot{z}_d$,

$$T_m = \dot{z}_e - \dot{z}_d.$$

Following the definitions used in [23],

$$\dot{z}_e = K_e \frac{Q}{d_s(1-\varphi)}(\theta - \theta_c)_+, \quad \dot{z}_d = K_d Q \frac{h_m}{d_s^2}, \tag{37}$$

$K_e$ and $K_d$ being the erosion and deposition constants respectively.

Nevertheless, it is necessary to remark that in such definition, the erosion rate does not take into account gravitational effects appearing in arbitrarily sloping beds. Thus, for the case of sediment beds which are not nearly horizontal, we propose to replace $\theta$ by $\theta_{\mathrm{eff}}$. Thus, we set

$$\dot{z}_e = K_e \frac{Q}{d_s(1-\varphi)}(\theta_{\mathrm{eff}} - \theta_c)_+$$

with $\theta_{\mathrm{eff}}$ defined by (25) or (33).

A simplification of the proposed model is to consider a quasi-uniform regime, where the deposition rate equals the erosion rate, that is $T_m = \dot{z}_e - \dot{z}_d = 0$. In this case, from the definition of $\dot{z}_d$ and $\dot{z}_e$, we have

$$h_m = \frac{K_e d_s}{K_d(1-\varphi)}(\theta_{\mathrm{eff}} - \theta_c)_+. \tag{38}$$

If we introduce this definition of $h_m$ in (17), we obtain the following system

$$\begin{cases} \partial_t h_1 + \mathrm{div}_x q_1 = 0, \\[2mm] \partial_t q_1 + \mathrm{div}_x(h_1(u_1 \otimes u_1)) + \dfrac{1}{2}g\nabla_x h_1^2 + gh_1\nabla_x(b + h_2) + \dfrac{gh_m}{r}\mathcal{P} = 0, \\[2mm] \partial_t h_2 + \mathrm{div}_x\left(\dfrac{K_e d_s}{K_d(1-\varphi)}(\theta_{\mathrm{eff}} - \theta_c)_+ v_b\,\sqrt{(1/r-1)gd_s}\right) = 0, \end{cases} \tag{39}$$

with $v_b$ and $\theta_{\mathrm{eff}}$ defined by (24)-(26) for $(LF)$ and by (30)-(33) for $(QF)$.

Let us focus on the case of a linearization of the quadratic friction law, for which $v_b$ is defined by (36). Then, we obtain a SVE model with the following definition of the solid transport discharge,

$$\frac{q_b}{Q} = \mathrm{sgn}(\tau_{\mathrm{eff}})\frac{K_e}{K_d(1-\varphi)}(\theta_{\mathrm{eff}} - \theta_c)_+ (\sqrt{\theta_{\mathrm{eff}}} - \sqrt{\theta_c}). \tag{40}$$

This definition can be seen as a generalization of the Ashida-Michiue model for arbitrarily sediment sloping beds, see equation (5).

Other possibilities for the definition of $h_m$ can be found in the bibliography. For instance, it is possible to define (see [56])

$$h_m = \frac{k_1 d_s}{(1-\varphi)} \frac{(\theta_{\text{eff}} - \theta_c)_+^{3/2}}{\sqrt{\theta_{\text{eff}}} - k_2 \sqrt{\theta_c}}, \tag{41}$$

$k_1$ and $k_2$ being two parameters of the model. Note that (41) is an approximation of (38), which is valid except near the threshold of motion.

Let us also remark that for the case $k_2 = 1$ and $v_b$ defined by (36) we obtain a SVE model with the following definition of the solid transport discharge,

$$\frac{q_b}{Q} = \text{sgn}(\tau_{\text{eff}}^{(QF)}) \frac{k_1}{(1-\varphi)} (\theta_{\text{eff}} - \theta_c)_+^{3/2}. \tag{42}$$

This definition can be seen as a generalization of the Meyer-Peter&Müller model for arbitrarily sediment sloping beds, see equation (4).

**Remark 2.3** *The simplified model defined by (42) coincides with the classical Meyer-Peter&Müller model for $k_1 = 8$ and by neglecting all gravitational terms in the definition of $\tau_{\text{eff}}$. That is, for $\tau_{\text{eff}}$ defined by (33) with $\Phi = \text{sgn}(u_1)\left(\vartheta \frac{d_s}{h_m} \theta^{(QF)}\right)^{1/2}$, where $\theta^{(QF)}$ is defined by (29).*

*We obtain also that with this definition of $\tau_{\text{eff}}$ the simplified model (40) coincides with the classical Ashida-Michiue model when $K_e/K_d = 17$.*

By analogous arguments we can deduce a large range of classical models. What allows us to see these classical models as simplified models which can been deduced from an asymptotic expansion of the Navier-Stokes equations, by considering the coupling between a shallow water layer and a Reynolds one. Then, for example – see previous remark–, we can deduce that Meyer-Peter&Müller model corresponds to define $h_m$ by (41) with $k_1 = 8$ and $k_2 = 1$. By while the Ashida-Michiue model corresponds to define $h_m$ by (38) with $K_e/K_d = 17$. In Figure 2 we can see the comparison of both definitions of $h_m$ for $\theta_c = 0.047$. Where we can observe that both definitions of $h_m$ are close.

## 2.3 Energy balance of the models

In this subsection we present the main result regarding the energy balance associated to the models $(LF)$ and $(QF)$ presented before. In particular we prove that they admit an exactly dissipation energy. The detailed proof of the theorem is given in the Appendix B.1.

**Theorem 2.1** *The model (17) has a dissipative energy balance. More explicitly:*

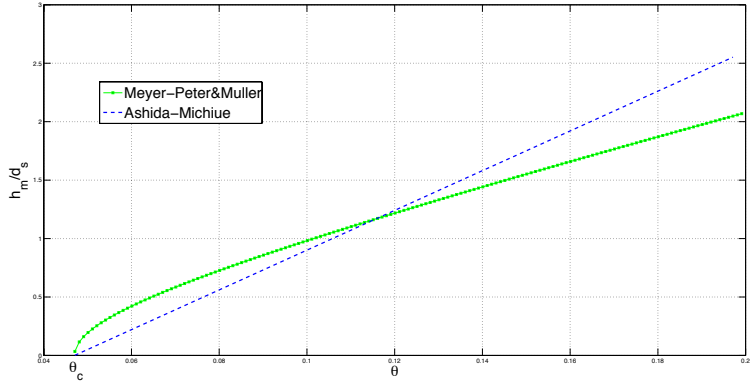

Figure 2: Definitions of $h_m$ corresponding to the Meyer-Peter&Müller and the Ashida-Michiue models.

- *for the (LF) model:*

$$
\begin{aligned}
&\partial_t \left( \frac{1}{2} r g (h_1 + h_2 + b)^2 + \frac{1}{2} r h_1 |u_1|^2 + g h_2 x (1-r) sgn(u_2) \tan \delta \right) \\
&+ div_x \left( r h_1 u_1 \left( \frac{|u_1|^2}{2} + g(h_1 + h_2 + b) \right) \right) \\
&+ div_x \left( h_m u_1 \tilde{\mathcal{P}} - h_m \frac{\vartheta}{1-r} \tilde{\mathcal{P}} \mathcal{P} \sqrt{(1/r - 1)g d_s} \right) \\
&\le -g h_m \frac{\vartheta}{1-r} \sqrt{(1/r - 1)g d_s} \, |\mathcal{P}|^2,
\end{aligned}
\tag{43}
$$

- *for the (QF) model:*

$$
\begin{aligned}
&\partial_t \left( \frac{1}{2} r g (h_1 + h_2 + b)^2 + \frac{1}{2} r h_1 |u_1|^2 + g h_2 x (1-r) sgn(u_2) \tan \delta \right) \\
&+ div_x \left( r h_1 u_1 \left( \frac{|u_1|^2}{2} + g(h_1 + h_2 + b) \right) \right) \\
&+ div_x \left( h_m u_1 \tilde{\mathcal{P}} - h_m \sqrt{\frac{\vartheta}{1-r}} \tilde{\mathcal{P}} |\mathcal{P}|^{1/2} sgn(\mathcal{P}) \sqrt{(1/r - 1)g d_s} \right) \\
&\le -g h_m \sqrt{\frac{\vartheta}{1-r}} \sqrt{(1/r - 1)g d_s} \, |\mathcal{P}|^{3/2},
\end{aligned}
\tag{44}
$$

*where* $\tilde{\mathcal{P}} = g\big((r h_1 + h_2 + b) + x(1-r) sgn(u_2) \tan \delta\big)$ *and* $\mathcal{P}$ *is defined in (18).*

## 2.4 Energy balance of the classical Saint-Venant-Exner model

As mentioned in the introduction there exist in the literature some simple solid transport formulae for which the corresponding SVE model has an associated dissipative energy equation. Nevertheless, up to our knowledge, no general result exists in the bibliography in this sense.

In this subsection we present a result that shows that by including a second order correction in the definition of the friction law in the momentum equation of the Saint-Venant system, we obtain a dissipative energy balance for any classical SVE model whose solid transport discharge can be written under the general form (2).

**Theorem 2.2** *Let us consider the general SVE system (1)-(3) and the following definition of the friction term in the momentum conservation equation of the Saint-Venant model,*

$$\tau/\rho_1 = g\zeta(h_1)u_1|u_1| + \frac{1}{r}\xi_m\sqrt{g\zeta(h_1)}\mathcal{R}.$$

*We have that if*

$$\sqrt{\theta} - \sqrt{\theta_c} + sgn(\tau)g\nabla_x(rh_1 + h_2 + b) > 0,$$

*then the SVE model satisfies the dissipative energy balance:*

$$\begin{aligned}
&\partial_t\left(\frac{1}{2}rg(h_1 + h_2 + b)^2 + \frac{1}{2}rh_1|u_1|^2 - h_2 sgn(\tau)\sqrt{\theta_c}\right)\\
&+ div_x\left(rh_1u_1\left(\frac{|u_1|^2}{2} + g(h_1 + h_2 + b)\right)\right)\\
&+ div_x\left(\xi_m u_1\sqrt{g\zeta(h_1)}\tilde{\mathcal{R}} - \xi_m\mathcal{R}\tilde{\mathcal{R}}\sqrt{(1/r - 1)gd_s}\right)\\
&\leq -g\zeta(h_1)|u_1|^3 - \xi_m\sqrt{(1/r - 1)gd_s}\,|\mathcal{R}|^2
\end{aligned} \tag{45}$$

*where*

$$\xi_m = \frac{1}{1 - \varphi}\,k_1\,\theta^{m_1}\,(\theta - k_2\theta_c)_+^{m_2}\,(\sqrt{\theta} - k_3\sqrt{\theta_c})_+^{m_3}\,\frac{1}{(\sqrt{\theta} - \sqrt{\theta_c} + sgn(\tau)g\nabla_x(rh_1 + h_2 + b))},$$

$$\mathcal{R} = sgn(\tau)\sqrt{\theta_c} - g\nabla_x(rh_1 + h_2 + b) \quad and \quad \tilde{\mathcal{R}} = sgn(\tau)x\sqrt{\theta_c} - g(rh_1 + h_2 + b).$$

The proof of this theorem is similar to the one of Theorem 2.1 and can be found in the Appendix B.2.

**Remark 2.4** *Note that the considered modification of the classical SVE model introduces a second order term in the system. That implies that the classical SVE model verifies a dissipative energy equation up to second order terms.*

# 3   Numerical tests

In this section we present three numerical tests, for the simplified model obtained from (17) with (41), that is, the solid transport equation reduces to (42). This coincides with a classical Meyer-Peter&Müller model with a modified shear stress given by $\tau_{\text{eff}}$ The purpose of these tests is to study the influence of the effective shear stress. The purpose is to study what are the effects of the different terms that appear in this modified shear stress.

The numerical results follow from a combination of the scheme described in [6] with a discrete approximation of bottom and surface derivatives. The numerical simulations are done with a CFL number equal to 0.5.

### 3.1 Test 1

In this first test we propose to study the influence of $\tan \delta = \theta_c/\vartheta$ in (24). To do so, let us consider the following initial condition

$$h_2(0, x) = \begin{cases} 0.2, & \text{if } x \in [4, 6], \\ 0.1, & \text{otherwise.} \end{cases} \qquad h_1(0, x) + h_2(0, x) = 1, \quad q_1(0, x) = 1.5. \qquad (46)$$

The initial condition is shown in Figure 3(a). We set the boundary condition $q(t, 0) = 1.5$.

The parameters for the model has been set as follows

$$r = 0.34, \quad d_s = 10^{-3},$$

$$\varphi = 0, \quad K_e = 10, \quad K_d = 1, \quad \theta_c = 0.047$$

with the Manning friction law and as Manning coefficient $n = 0.01$. The computational domain used is $[0, 10]$ with 800 points.

Remark that this is a rather severe test: the bottom is discontinuous and thus we have $\partial_x(h_2 + b) = \pm\infty$ initially. We run the test for different values of $\delta$ ranging from $25°$ to $89°$. In Figure 3(b) we show the surface and bottom at time $t = 2000$ for $\delta = 89°$, $60°$, $45°$ and $25°$. The results are also compared with a classical Meyer-Peter&Müller model. A more detailed comparison of the final bottom can be seen in Figure 4(a)

Remark that for $\delta \to 90°$ we have $\kappa \to 0$. Thus, the model reduces to a classical Meyer-Peter&Müller model for values of $\delta$ near $90°$ as it can be observed in Figure 3(c). For smaller values of $\delta$, the effects of the gradient $\partial_x(rh_1 + \eta)$ play a relevent role in the stress tensor. This is shown, for instance, in Figures 4(b) and 4(c), where the value $\tau - \tau_{eff}$ displayed. The steep profile of $\partial_x(rh_1 + \eta)$ makes that this term plays an esentiall role in the evolution of the test. The smaller the value $\delta$, the bigger the influence of the gradient on this test. This is specially true for small values of $t$. This results in a smoothing of the dune profile, and in particular in the advancing front of the dune.

### 3.2 Test 2

The purpose of this test is to study the influence on the definition of the effective shear stress in terms of $\nabla_x \eta$ instead of $\nabla_x(rh_1 + \eta)/(1 - r)$. Following Remark 2.1, both terms can be related by

$$\nabla_x \eta \quad = \frac{1}{1 - r} \nabla_x(rh_1 + \eta) - \frac{r}{1 - r} \nabla_x(b + h_1 + h_2). \qquad (47)$$

This means that both definitions coincide only for constant free surface. We recall that the model proposed in [26] defines the effective shear in terms of $\nabla_x \eta$. This model was also studied in [40].

We set as initial condition

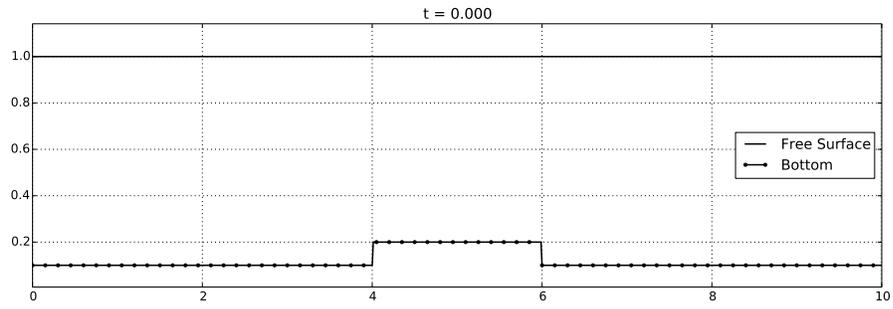

(a) Test 1: Initial condition

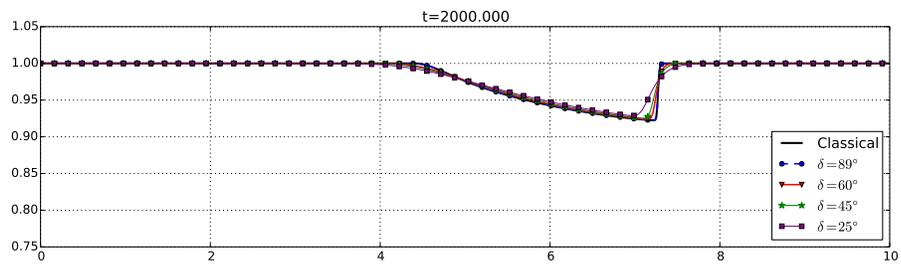

(b) Test 1: Surface at time $t = 2000$

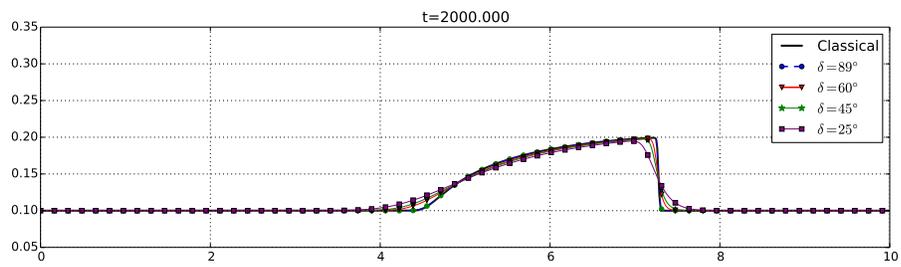

(c) Test 1: Bottom at time $t = 2000$

Figure 3: Test 1: Initial condition and evolution for different values of $\delta$

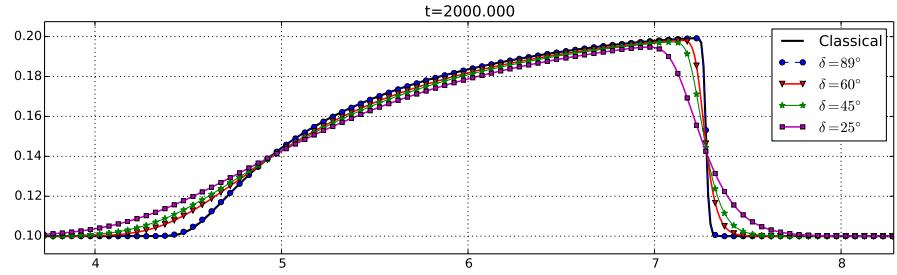

(a) Test 1: Bottom comparison (Zoom)

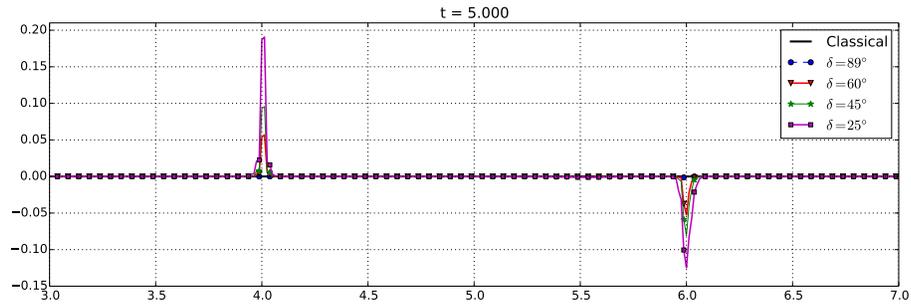

(b) Test 1: $\tau - \tau_{eff}$ at time $t = 5$ (Zoom)

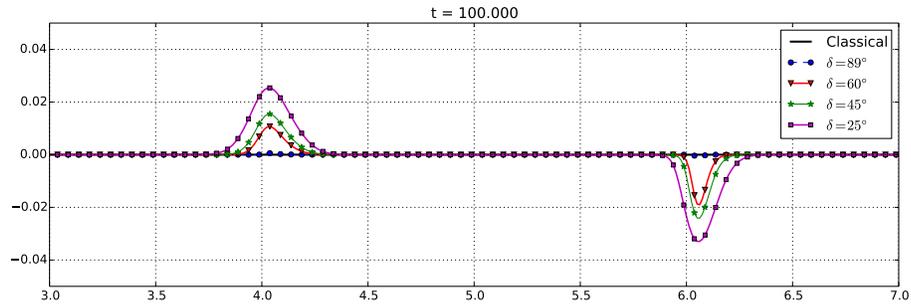

(c) Test 1: $\tau - \tau_{eff}$ at time $t = 100$ (Zoom)

Figure 4: Test 1: Influence of $\delta$

$$h_2(0, x) = \begin{cases} 0.1 + 0.1 \left(1 + \cos\left(\dfrac{x - 0.4}{0.2\pi}\right)\right), & \text{if } x \in [0.2, 0.6], \\ 0.1, & \text{otherwise.} \end{cases} \tag{48}$$

$$h_1(0, x) + h_2(0, x) = 1, \quad q_1(0, x) = 1.4, \tag{49}$$

The initial condition is shown in Figure 5.

We set the boundary condition $q(t, 0) = 1.4$ and we have used the same parameters and computational domain described int Test 1. $\delta$ is fixed to $45°$.

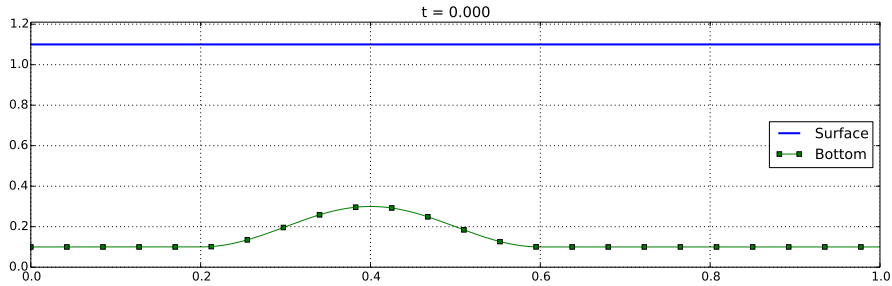

Figure 5: Test 2: Initial condition

Figures 6(a) and 6(b) show the difference observed on the surface and bottom when the different terms $\partial_x(rh_1 + \eta)$ and $\partial_x\eta$ are used in the definition of $\tau_{eff}$. Due to the shape of the surface and bottom, we get that $\partial_x\eta \geq \frac{1}{1-r}(rh_1 + \eta)$ and $\partial_x\eta \leq \frac{1}{1-r}(rh_1 + \eta)$ on the upstream and downstream part of the dune respectively which is shown in 6(c). This makes that the influence of the gravitational effects in $\tau_{eff}$ are stronger when $\partial_x\eta$ is used and results as a more difussed shape observed in Figure 6(b).

## 3.3 Test 3

In this section we present a comparison with experimental data obtained by the Hydraulic Laboratory of Escuela Superior de Ingenieros de Caminos, Canales y Puertos (A Coruña University) over a channel of 15 $m$ long and 0.5 $m$ width (for more details see [49] and [6]). We compare the experimental data with the numerical simulation corresponding to the modified Meyer-Peter&Müler's model, defined by (42).

The experimental test was developed by introducing a sand layer in the central part of laboratory channel, and inducing hydrodynamical conditions to erode the sand layer. The channel has slope of 0.052 %. Sand layer was situated in interval $[4.5\ m, 9\ m]$, with a thickness of 4.5 $cm$; being media diameter of the grain equals to 1 $mm$. As boundary conditions, an incoming discharge equal to 0.0285 $m^2/s$ upstream is imposed. The water thickness is 0.129 $m$ downstream.

We have the experimental measurements of the sediment profile at several points for $t = 10$, $t = 40$ and $t = 120$ minutes. Then, for the numerical simulation we have considered

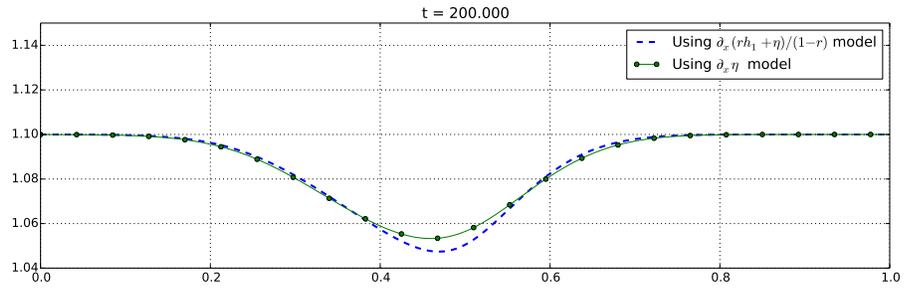

(a) Test 2: Surface at time $t = 200$

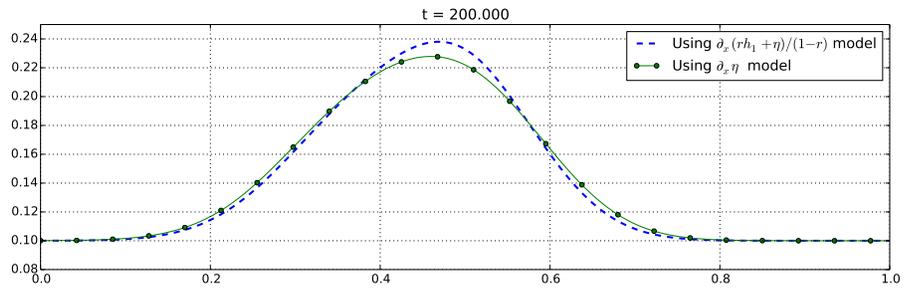

(b) Test 2: Bottom at time $t = 200$

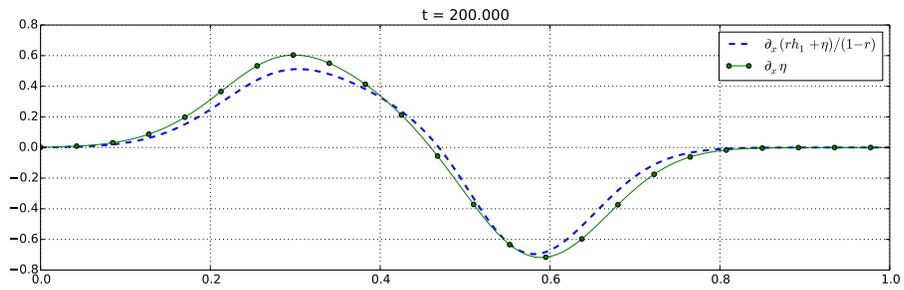

(c) Test 2: $\partial_x(rh_1 + \eta)$ and $\partial_x\eta$ at time $t = 200$

Figure 6: Test 2: Comparisons at time $t = 200$

as initial condition for the sediment surface a profile produced by an interpolation of the data at $t = 10$ minutes (see Figure 7). The initial condition for the free surface and the discharge can be precomputed by considering this profile of the sediment layer and previous boundary conditions.

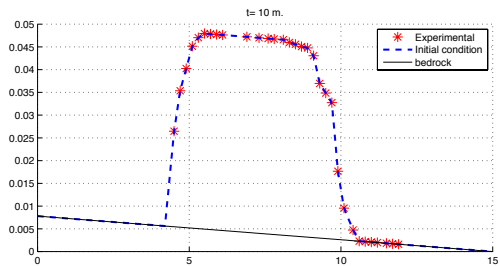

Figure 7: Test 3: Sediment surface. Initial condition

For numerical simulation we have meshed the domain with 250 nodes. The CFL is set to 0.9. Sediment porosity is set to 0.4. Friction between fluid and bed is modeled using a Manning's law with coefficient equal to 0.0125 over the fixed bed and 0.0196 over the sediment layer.

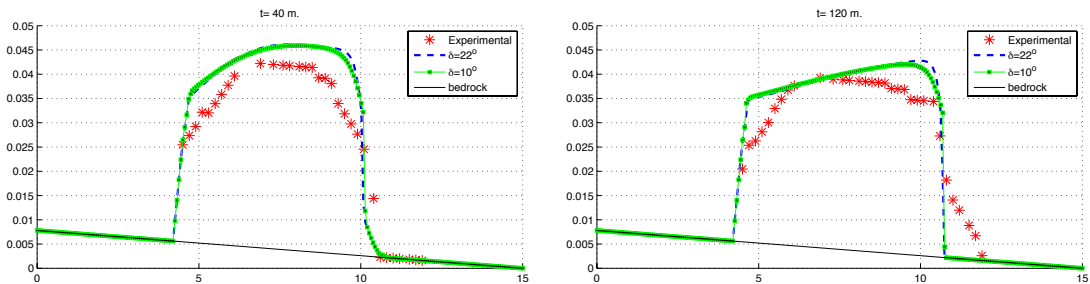

Figure 8: Test 3: Sediment surface evolution at $t = 40$ min. and $t = 120$ min.

Comparisons with the experimental data for $t = 40$ and $t = 120$ minutes are presented in Figure 8. Two numerical simulations has been done, corresponding to the Coulomb friction angle $\delta = 10^o$ and $\delta = 22^o$. We can observe that in both cases the averaged position of the sediment surface is well reproduced by the numerical simulations. We can remark that at $t = 120$ minutes the numerical simulation reproduces correctly the position of the discontinuity in the profile of the sediment bed, located at $x \approx 10.7$. Nevertheless, the model does not produce the small sediment bed at the right of the shock. Although, we have tested a large range of models to define the solid transport discharge and any of them reproduce this advance of the sediment bed at the right of the shock. Which is probably a purely tridimensional effect.

# 4 Conclusions

The SVE model has been deduced in this work by an asymptotic analysis of the Navier-Stokes equations. We show that depending on the considered friction law at the interface between the fluid and the sediment we can obtain different definitions of solid transport discharge. Results have been presented for the case of linear and quadratic friction laws. The stop criteria, which is usually represented as a positive part in terms of a critical Shields stress, is deduced by including a Coulomb friction law in the internal interface of the sediment layer.

The proposed models present also some advantages with respect to classical ones: they have an associated dissipative energy, they are deduced for arbitrarily sediment sloping beds and they depend on the thickness of the moving sediment layer, what implies sediment mass conservation in general situations.

Moreover, from the deduction of the model we obtain a modification on the definition of the effective shear stress. We obtain that it must be defined in terms of the gradient of total pressure, $\nabla_x(rh_1 + b + h_2)$. By while, classical models consider the gradient of sediment bed, $\nabla_x(b + h_2)$. Nevertheless, by considering this classical definition we cannot obtain a model with a dissipative energy.

Finally, the deduction of the model and the proof of the existing energy dissipation equation provides important information on classical models. It allows us to understand why classical SVE models do not have an associated energy. Moreover, in order to have an associated dissipative energy we may we add an extra term – a term that is deduced in the modelling process – in the definition of the shear stress which appears in the momentum equation of the Saint-Venant model. This allows also us to deduce that any classical SVE exner model, defined by a solid transport discharge which can be written under the general form (2), verifies a dissipative energy equation up to second order.

In the numerical test we presented the evolution of a dune for several repose angle. We can see as the morphodynamical form of the dune is well reproduced. The influence of the definition of the effective shear stress as we propose in this paper in comparison with the classical one is compared in the second test. We can observe that a difference appears between both definitions just in the area over the dune, where the free surface is not constant. In the last numerical test we have compared with experimental data for the generalization of the Meyer-Peter&-Müller model. We obtain that the averaged position of the dune is well captured in comparison with experimental data. And the position of the shock is well approximated.

## Acknowledgments

This research has been partially supported by the Spanish Government and FEDER through the Research project MTM2012-38383-C02-02, and by the Andalusian Government through the projects P11- RNM7069 and P11-FQM8179.

Authors would like to thank to Pierre-Yves Lagrée for his comments and the interesting discussions that followed during his visit to the University of Seville in November 2014.

# Appendix

# A    Derivation of the sediment transport models

In this section we develop the derivation of the sediment transport models from an asymptotic analysis of the Navier-Stokes equations. These processes are different for each layer. In [24] a similar development is performed coupling also shallow water and Reynolds equations in a multiscale framework but for a transport of pollutant problem. Thus, we consider this work as a reference in the development, which in turn is based on the original works [30] and [44].

The derivation is divided in several steps: first, we write the initial system in non-dimensional form; next we make an hydrostatic approximation, we assume a suitable asymptotic regime and we average the equations out.

## A.1    Dimensionless equations

To begin with, we write the equations and boundary conditions under dimensionless form. We set the dimensionless variables, where we must take into account the different nature of the fluids in two layers, so we make it separately. Indeed the main property that we want to point out is the different order of the velocities, as we have discussed in Section 2.1. Since we study a coupled system we relate the two characteristic velocities by the aspect ratio to indicate that the sediment layer is slower than the fluid layer.

We denote by $H$ and $L$ the characteristic height and length respectively. To impose the shallow flow condition, we assume that the aspect ratio between the characteristic height and length is small, as usual we denote it by $\varepsilon = \frac{H}{L}$. The characteristic velocities are $U$ for the layer 1 and $U_2$ for the sediment layer, consequently, the characteristic times are respectively $T = \frac{L}{U}$ and $T_2 = \frac{L}{U_2}$ for each layer. This hypothesis also affects the definitions of the Froude and Reynolds numbers. For the sake of clarity we indicate separately these variables.

We consider the "star" notation for the dimensionless variables.

*General dimensionless variables:*

$$x = Lx^* \qquad\qquad z = Hz^* \qquad\qquad h_i = Hh_i^* \quad b = Hb^*$$

$$\mathrm{fric}_f = \rho_2 U^2 \mathrm{fric}_f^* \quad \mathrm{fric} = \rho_1 U^2 \mathrm{fric}_H^*$$

*Nondimensionalization for layer 1:*

$$u_1 = Uu_1^* \quad w_1 = \varepsilon Uw_1^* \quad t = \frac{L}{U}t_1^* \quad p_1 = \rho_1 U^2 p_1^*$$

$$Re_1 = \frac{UL}{\nu_1} \quad Fr_1 = \frac{U}{\sqrt{gH}} \quad \mu_1 = \rho_1 \nu_1$$

*Nondimensionalization for layer 2:*

$$u_2 = U_2 u_2^* \quad w_2 = \varepsilon U_2 w_2^* \quad t = \frac{L}{U_2} t_2^* \quad p_2 = \frac{\rho_2 \nu_2 U_2}{\varepsilon H} p_2^*$$

$$Re_2 = \frac{U_2 L}{\nu_2} \quad Fr_2 = \frac{U_2}{\sqrt{gH}} \quad \mu_2 = \rho_2 \nu_2 \mu_2^* \quad T_m = \varepsilon U_2 T_m^*$$

Since we study the coupled system we must take into account that layer 2 is slower than layer 1. This related aspect lead us to search for a relationship between the characteristic velocities of the two layers holding this property. Thus, we assume that

$$U_2 = \varepsilon^2 U,$$

so consequently, $T_2 = \frac{L}{\varepsilon^2 U} = \frac{1}{\varepsilon^2} T$ and we can write the dimensionless variables for the sediment layer in terms of $\varepsilon$ as:

$$u_2 = \varepsilon^2 U u_2^* \quad w_2 = \varepsilon^3 U w_2^* \quad t = \frac{L}{\varepsilon^2 U} t_2^* \quad p_2 = \frac{\varepsilon \rho_2 \nu_2 U}{H} p_2^*$$

$$Re_2 = \varepsilon^2 \frac{UL}{\nu_2} \quad Fr_2 = \varepsilon^2 \frac{U}{\sqrt{gH}} \quad \mu_2 = \rho_2 \nu_2 \mu_2^* \quad T_m = \varepsilon^3 U T_m^*$$

We also define the ratio of densities,

$$r = \frac{\rho_1}{\rho_2} < 1.$$

Thus, the equations and the boundary conditions written in dimensionless form read as follows (we omit the "star" to simplify the notation):

$$\begin{cases} \partial_{t_1} u_1 + \text{div}_x(u_1 \otimes u_1) + \partial_z(u_1 w_1) - \frac{2}{Re_1} \text{div}_x(D_x u_1) \\[4pt] \quad - \frac{1}{\varepsilon^2} \frac{1}{Re_1} \partial_z^2 u_1 - \frac{1}{Re_1} \nabla_x(\partial_z w_1) + \nabla_x p_1 = 0; \\[10pt] \partial_{t_1} w_1 + u_1 \nabla_x w_1 + w_1 \partial_z w_1 - \frac{1}{Re_1} \text{div}_x(\nabla_x w_1) \\[4pt] \quad - \frac{1}{\varepsilon^2} \frac{1}{Re_1} \partial_z(\text{div}_x u_1) - 2\frac{1}{\varepsilon^2} \frac{1}{Re_1} \partial_z^2 w_1 + \frac{1}{\varepsilon^2} \partial_z p_1 = -\frac{1}{\varepsilon^2} \frac{1}{Fr_1^2}; \\[10pt] \text{div}_x u_1 + \partial_z w_1 = 0. \end{cases} \tag{50}$$

$$\begin{cases} \varepsilon^2 Re_2(\partial_{t_2} u_2 + \text{div}_x(u_2 \otimes u_2) + \partial_z(u_2 w_2)) \\[4pt] \quad = -\nabla_x p_2 + \partial_z(\mu_2 \partial_z u_2) + 2\varepsilon^2 \text{div}_x(\mu_2 D_x u_2) + \varepsilon^2 \nabla_x(\mu_2 \partial_z w_2); \\[10pt] \varepsilon^4 Re_2(\partial_{t_2} w_2 + u_2 \nabla_x w_2 + w_2 \partial_z w_2) \\[4pt] \quad = -\partial_z p_2 + \varepsilon^2(2\partial_z(\mu_2 \partial_z w_2) + \partial_z(\mu_2 \text{div}_x u_2) + \varepsilon^2 \text{div}_x(\mu_2 \nabla_x w_2)) - \varepsilon^2 \frac{Re_2}{Fr_2^2}; \\[10pt] \text{div}_x u_2 + \partial_z w_2 = 0. \end{cases} \tag{51}$$

- Conditions at the free surface:

$$\partial_{t_1}\eta_s + u_1 \cdot \nabla_x \eta_s = w_1; \tag{52}$$

$$\left(-\frac{2}{Re_1}D_x u_1 + p_1\right)\partial_x \eta_s + \frac{1}{Re_1}\nabla_x w_1 + \frac{1}{\varepsilon^2}\frac{1}{Re_1}\partial_z u_1 = 0; \tag{53}$$

$$-\frac{1}{Re_1}\left(\varepsilon^2\nabla_x w_1 + \partial_z u_1\right)\nabla_x \eta_s + \frac{2}{Re_1}\partial_z w_1 - p_1 = 0. \tag{54}$$

- Conditions at the interface:

$$\partial_{t_1}\eta + u_1 \cdot \nabla_x \eta = w_1; \tag{55}$$

$$\partial_{t_2}\eta + u_2 \cdot \nabla_x \eta = w_2; \tag{56}$$

$$r\varepsilon^2|\nabla\eta|^2\left(2\frac{1}{Re_1}D_x u_1 - p_1\right) + r\frac{1}{Re_1}\left(-2(\varepsilon^2\nabla_x w_1 + \partial_z u_1)\nabla_x\eta + 2\partial_z w_1\right) - rp_1$$

$$= -\frac{1}{Re_2}\left(\varepsilon^6|\nabla_x\eta|^2(2\mu_2 D_x u_2 - p_2) - 2\varepsilon^4\mu_2(\varepsilon^2\nabla_x w_2 + \partial_z u_2)\nabla_x\eta + 2\varepsilon^4\mu_2\partial_z w_2 - \varepsilon^2 p_2\right); \tag{57}$$

$$\frac{1}{Re_1}\left(-2\nabla_x\eta(D_x u_1 - \partial_z w_1) + (\nabla_x w_1 + \frac{1}{\varepsilon^2}\partial_z u_1)(1 - \varepsilon^2|\nabla_x\eta|^2)\right)$$
$$= \frac{1}{\varepsilon}\text{fric}\sqrt{1 + \varepsilon^2|\nabla_x\eta|^2}; \tag{58}$$

$$\varepsilon^4\mu_2\frac{1}{Re_2}\left(-2\nabla_x\eta(D_x u_2 - \partial_z w_2) + (\nabla_x w_2 + \frac{1}{\varepsilon^2}\partial_z u_2)(1 - \varepsilon^2|\nabla_x\eta|^2)\right)$$
$$= r\frac{1}{\varepsilon}\text{fric}\sqrt{1 + \varepsilon^2|\nabla_x\eta|^2}; \tag{59}$$

- Conditions at the internal sediment interface:

$$\partial_{t_2}h_f + u_2\nabla_x\eta_f - w_2 = -T_m; \tag{60}$$

$$\varepsilon^4\mu_2\frac{1}{Re_2}\left(-2\nabla_x\eta(D_x u_2 - \partial_z w_2) + (\nabla_x w_2 + \frac{1}{\varepsilon^2}\partial_z u_2)(1 - \varepsilon^2|\nabla_x\eta|^2)\right)$$
$$= \frac{1}{\varepsilon}\text{fric}_f\sqrt{1 + \varepsilon^2|\nabla_x\eta|^2}; \tag{61}$$

- Condition at the bottom:

$$-u_2\nabla_x b + w_2 = 0. \tag{62}$$

## A.2 Layer $\Omega_1$: shallow water.

In this section we obtain the mass and momentum approximated equations for the fluid layer.

**Hydrostatic approximation**

We assume $\varepsilon$ to be small and keep the terms in the system of order zero and one. Then we integrate this system in $[\eta, \eta_s]$. First, we integrate the equation of the horizontal velocity and use conditions (52), (53), (54), (55) and (58) to obtain:

$$\partial_{t_1}\left(\int_\eta^{\eta_s} u_1 dz\right) + \text{div}_x\left(\int_\eta^{\eta_s}(u_1 \otimes u_1)dz\right) - \frac{2}{Re_1}\text{div}_x\left(\int_\eta^{\eta_s} D_x u_1 dz\right) + \nabla_x\left(\int_\eta^{\eta_s} p_1 dz\right)$$
$$- \frac{2}{Re_1}\partial_z w_{1|z=\eta}\nabla_x\eta + p_{1|z=\eta}\nabla_x\eta + \frac{1}{\varepsilon}\text{fric} = \mathcal{O}(\varepsilon^2). \tag{63}$$

Now, to get $p_1$ we integrate the vertical velocity equation from $z$ to $\eta_s$ and use the divergence free condition:

$$p_1(z) = p_{1|z=\eta_s} - \frac{1}{Re_1}(\text{div}_x u_1 - \text{div}_x u_{1|z=\eta_s}) - \frac{1}{Fr_1^2}(z - \eta_s) + \mathcal{O}(\varepsilon^2). \tag{64}$$

**Asymptotic analysis**

To develop the asymptotic analysis, we assume the following hypotheses on the data:

$$\frac{1}{Re_1} = \varepsilon\nu_{01}, \quad \text{fric} = \varepsilon\,\text{fric}_0,$$

and we consider the development of the unknowns in terms of $\varepsilon$:

$$h_1 = h_1^0 + \varepsilon h_1^1 + \mathcal{O}(\varepsilon^2), \quad u_1 = u_1^0 + \varepsilon u_1^1 + \mathcal{O}(\varepsilon^2), \quad p_1 = p_1^0 + \varepsilon p_1^1 + \mathcal{O}(\varepsilon^2).$$

We also introduce for simplicity the notation $\eta^0$ and $\eta_s^0$ to write the main order components of the interface and the free surface respectively.

If we just consider the terms of the principal order ($\varepsilon^0$), we obtain from first equation in (50), (58) and (53) that:

$$\partial_z^2 u_1 = \mathcal{O}(\varepsilon);$$
$$\partial_z u_{1|z=\eta} = \mathcal{O}(\varepsilon);$$
$$\partial_z u_{1|z=\eta_s} = \mathcal{O}(\varepsilon),$$

from where we deduce that $u_1$ does not depend on $z$ at first order, so:

$$u_1^0(t, x, z) = u_1^0(t, x).$$

This is the usual "motion by slices" property of the shallow flows. Under this hypothesis, we shall rewrite the expressions above to obtain the final equation for layer 1 at first order. First, we write the mass conservation equation from (50) as:

$$\partial_t h_1^0 + \text{div}_x(h_1^0 u_1^0) = 0. \tag{65}$$

From (64) and taking into account the free surface condition (54), the pressure reads,

$$p_1(z) = -\frac{1}{Fr_1^2}(z - \eta_s^0) - 2\varepsilon\nu_{01}\mathrm{div}_x u_1^0 + \mathcal{O}(\varepsilon^2). \tag{66}$$

Using these values in equation (63), we obtain:

$$\partial_{t_1}(h_1^0 u_1^0) + \mathrm{div}_x(h_1^0(u_1^0 \otimes u_1^0)) + \frac{1}{2}\frac{1}{Fr_1^2}\nabla_x(h_1^0)^2 + \frac{1}{Fr_1^2}h_1^0\nabla_x\eta^0 + \mathrm{fric}_0 = \mathcal{O}(\varepsilon), \tag{67}$$

where the friction term $\mathrm{fric}_0$ will be specified later.

## A.3  Layer $\Omega_2$: Reynolds.

We derive in this section the first order approximation of the evolution equation for the layer 2. Due to the more complex dimensionless considered for this layer, it is necessary to keep in the derivation process the terms of order zero and one (the unknowns are denoted with tilde). Later we will only consider the terms of principal order.

We write the momentum equations in (51) up to second order :

$$-\partial_z(\mu_2\partial_z u_2) + \nabla_x p_2 = \mathcal{O}(\varepsilon^2); \tag{68}$$

$$\partial_z p_2 + \varepsilon^2\frac{Re_2}{Fr_2^2} = \mathcal{O}(\varepsilon^2). \tag{69}$$

The mass equation for the sediment layer comes from the integration of the incompressibility equation. Since we decomposed this layer into the static and the moving sediment layers, we make this integration in each of them. Thus, for the static layer we have that $u_2(z) = 0$ for $b \leq z < \eta_f$, and we use conditions (60) and (62) to get

$$\partial_{t_2}h_f = -T_m \tag{70}$$

Using conditions (56) and (60) we write the mass conservation for the moving layer:

$$\partial_{t_2}h_m + \mathrm{div}_x\left(\int_{\eta_f}^{\eta} u_2 dz\right) = T_m. \tag{71}$$

The expression for the velocity $u_2$ will be obtained from the equation (68), but first we need to know the value of the pressure $p_2$ that will be given by equation (69).

**Remark A.1** *Due to the incompressibility equation, we only need to find an equation for $u_2$. In fact from the mass conservation equation, we have*

$$w_2(z) = w_{2|z=\eta} - \int_z^{\eta} div_x u_2 dz,$$

*where $w_{2|z=\eta}$ is given by the kinematic condition at the interface (56).*  □

**Asymptotic analysis**

We consider the following asymptotic regime in order to keep the gravitational and inter-granular friction effects:

$$\nu_2 = \varepsilon^{-1}\bar{\nu}_2; \quad \text{fric}_f = \varepsilon\,\text{fric}_{f_0};$$

Thanks to the definition of the dimensionless variables for the layer 2, we have $Re_2 = \frac{\varepsilon^2 UL}{\nu_2}$, so for simplicity we will use the notation

$$Re_2 = \frac{\varepsilon^3}{\nu_{02}}, \quad \text{where } \nu_{02} = \frac{\bar{\nu}_2}{UL} = \mathcal{O}(1).$$

Notice that from the definitions of $Re_2$ and $Fr_2$ we have $\varepsilon^2\frac{Re_2}{Fr_2^2} = \frac{gLH}{\nu_2 U}$, so for simplicity we introduce

$$\beta_0 = \varepsilon\frac{Re_2}{Fr_2^2} = \varepsilon\frac{1}{\nu_{02}Fr_1^2}. \tag{72}$$

As we did for the layer 1, we develop each unknown in terms of $\varepsilon$ and introduce the notation:

$$\tilde{h}_m = h_m^0 + \varepsilon h_m^1, \quad \tilde{u}_2 = u_2^0 + \varepsilon u_2^1, \quad \tilde{p}_2 = p_2^0 + \varepsilon p_2^1.$$

Note that the integration at this level involves only the moving layer $h_m$, because the velocity is assumed to be zero for $b \leq z < \eta_f$.

So to obtain the pressure we integrate the equation (69) for $z \in [\eta_f, \eta]$:

$$\tilde{p}_2(z) = \tilde{p}_{2|z=\eta} - \varepsilon\beta_0(z - \eta^0). \tag{73}$$

We use the interface condition (57) to get the value of the pressure at the interface up to second order

$$p_{2|z=\eta} = \varepsilon\frac{r}{\nu_{02}}(p_{1|z=\eta} + 2\varepsilon\nu_{01}\text{div}_x u_1^0) + \mathcal{O}(\varepsilon^2). \tag{74}$$

Thanks to (66) we get:

$$\tilde{p}_2(z) = \varepsilon\frac{r}{\nu_{02}}\frac{1}{Fr_1^2}h_1^0 - \varepsilon\beta_0(z - \eta^0). \tag{75}$$

Thus

$$\nabla_x\tilde{p}_2 = \varepsilon\frac{r}{\nu_{02}}\frac{1}{Fr_1^2}\nabla_x h_1^0 + \varepsilon\beta_0\nabla_x\eta^0, \tag{76}$$

that does not depend on $z$. Taking into account the definition of $\beta_0$, we can write this equation as follows:

$$\nabla_x\tilde{p}_2 = \varepsilon\frac{1}{\nu_{02}}\frac{1}{Fr_1^2}(r\nabla_x h_1^0 + \nabla_x\eta^0). \tag{77}$$

Next in order to get an expression for the velocity $\tilde{u}_2$, we integrate (68) from $\eta_f$ to $z$ to find

$$\mu_2\partial_z\tilde{u}_2 = (\mu_2\partial_z\tilde{u}_2)_{|z=\eta_f} + (z - \eta_f)\nabla_x\tilde{p}_2.$$

We get the value at the internal sediment interface from the friction condition (61) at second order:

$$\nu_{02}(\mu_2 \partial_z u_2)_{|z=\eta_f} = \varepsilon \text{fric}_{f_0} + \mathcal{O}(\varepsilon^2).$$ (78)

Then we obtain

$$\partial_z \tilde{u}_2 = \frac{1}{\mu_2} \frac{\varepsilon}{\nu_{02}} \text{fric}_{f_0} + \nabla_x \tilde{p}_2 \frac{z - \eta_f}{\mu_2}.$$ (79)

We integrate again but now from $z$ to $\eta$ to get $\tilde{u}_2$. We must take into account that $\mu_2$ is not constant, so

$$\tilde{u}_2(z) = \tilde{u}_{2|z=\eta} + \frac{\varepsilon}{\nu_{02}} \text{fric}_{f_0} \int_\eta^z \frac{1}{\mu_2} dz + \nabla_x \tilde{p}_2 \int_\eta^z \frac{z - \eta_f}{\mu_2} dz.$$ (80)

Since we are interested in the principal order approximation, we neglect the two last terms —remember that $\nabla_x \tilde{p}_2 \sim \mathcal{O}(\varepsilon)$ from equation (77)—. So we approximate

$$\tilde{u}_2(z) = \tilde{u}_{2|z=\eta} + \mathcal{O}(\varepsilon).$$

If we use the boundary condition at the interface (59),

$$(\mu_2 \partial_z \tilde{u}_2)_{|z=\eta} = \varepsilon \frac{r}{\nu_{02}} \text{fric}_0 + \mathcal{O}(\varepsilon^2)$$ (81)

and the previous expression for $\mu_2 \partial_z \tilde{u}_2$ evaluated in $z = \eta$ we get a relation between the friction forces:

$$r \, \text{fric}_0 = \text{fric}_{f_0} + \frac{h_m}{Fr_1^2} (r \nabla_x h_1^0 + \nabla_x \eta^0)$$ (82)

where we have used (77).

At this moment, in order to find an expression for $\tilde{u}_{2|z=\eta}$, and then for $\tilde{u}_2$, we must explicit the friction terms.

For the friction at the level $z = \eta_f$, we consider a Coulomb friction law,

$$\text{fric}_f = -\Big(\text{sgn}(u_2) \tan \delta \big((\sigma_1 - \sigma_2) N_f\big) \cdot N_f\Big)_{|z=\eta_f}$$ (83)

with $N_f$ the unitary normal vector to the interface $z = \eta_f$ and $\delta$ the intergranular Coulomb friction angle. To be consistent with the development done before, the asymptotic assumption must be

$$\tan \delta = \varepsilon \tan \delta_0.$$

Developing this expression and using (66) we have:

$$\text{fric}_{f_0} = -\frac{r-1}{Fr_1^2} \text{sgn}(u_2) \tan \delta_0 \tilde{h}_m.$$ (84)

We consider two possible friction laws for the level $z = \eta$ and we derive the corresponding models.

- **Linear friction law at** $z = \eta$

We use a generalized law based in the work [42] (see Remark A.2), that reads

$$\text{fric} = C(u_1 - u_2)_{|z=\eta} \tag{85}$$

**Remark A.2** *The friction law introduced in (85) is based on the work performed in [42]. If we take into account the asymptotic regime considered for the viscosities, the coefficient $C$ can be written (up to second order) as:*

$$C = k h_m.$$

*We must also take care of the adimensionalization for this friction term. Thus we assume the following dimension and asymptotic to the coefficient $C$:*

$$C = U C^*; \quad C^* = \varepsilon C^0.$$

Then we have

$$\text{fric}_0 = C^0 \left( u_1^0 - \varepsilon^2 \tilde{u}_{2|z=\eta} \right) \tag{86}$$

From the friction at $z = \eta$ (86) and using (82) we get the value of $\tilde{u}_{2|z=\eta}$

$$
\begin{aligned}
u_{2|z=\eta} &= \frac{1}{\varepsilon^2} u_1^0 - \frac{1}{\varepsilon^2 C^0} \text{fric}_0 \\
&= \frac{1}{\varepsilon^2} u_1^0 - \frac{1}{r \, \varepsilon^2 C^0} \left( \text{fric}_{f_0} + \frac{h_m}{Fr_1^2} (r \nabla_x h_1^0 + \nabla_x \eta^0) \right)
\end{aligned}
$$

Finally, an expression for $\tilde{u}_2$ is deduced:

$$\tilde{u}_2(z) = \frac{1}{\varepsilon^2} u_1^0 - \frac{1}{r \, \varepsilon^2 C^0} \frac{1}{Fr_1^2} \tilde{h}_m \left( (1-r)\text{sgn}(u_2) \tan \delta_0 + (r \nabla_x h_1^0 + \nabla_x \eta^0) \right) \tag{87}$$

Now we use the relation (71) to write an equation for $h_m$ up to first order:

$$\partial_{t_2} h_m^0 + \text{div}_x \left( \frac{1}{\varepsilon^2} h_m^0 u_1^0 - \frac{1}{r \, \varepsilon^2 C^0} \frac{(h_m^0)^2}{Fr_1^2} \left( (1-r)\text{sgn}(u_2) \tan \delta_0 + (r \nabla_x h_1^0 + \nabla_x \eta^0) \right) \right) = T_m. \tag{88}$$

Finally we use (82) to complete the equation for the momentum of the layer 1, (67) that reads:

$$
\begin{aligned}
&\partial_{t_1} (h_1^0 u_1^0) + \text{div}_x (h_1^0 (u_1^0 \otimes u_1^0)) + \tfrac{1}{2} \tfrac{1}{Fr_1^2} \nabla_x (h_1^0)^2 + \frac{1}{Fr_1^2} h_1^0 \nabla_x \eta^0 \\
&+ \frac{1}{r} \frac{1}{Fr_1^2} \tilde{h}_m \left( (1-r)\text{sgn}(u_2) \tan \delta_0 + (r \nabla_x h_1^0 + \nabla_x \eta^0) \right) = \mathcal{O}(\varepsilon).
\end{aligned}
\tag{89}
$$

The final model is given by equations (65), (70), (88) and (89). The dimension form of the system is given by (17)-(19).

- **Quadratic friction law at** $z = \eta$

We consider now a quadratic law as follows

$$\text{fric} = C_1|(u_1 - u_2)_{|z=\eta}|(u_1 - u_2)_{|z=\eta} \tag{90}$$

with $C_1$ an adimensional constant that we suppose of order $\varepsilon$, that is, $C_1 = \varepsilon C_1^0$. Then,

$$\text{fric}_0 = C_1^0 \left| u_1^0 - \varepsilon^2 \tilde{u}_{2|z=\eta} \right| (u_1^0 - \varepsilon^2 \tilde{u}_{2|z=\eta}).$$

We can find a solution for this equation, given by:

$$u_1^0 - \varepsilon^2 \tilde{u}_{2|z=\eta} = \frac{1}{\sqrt{C_1^0}} |\text{fric}_0|^{1/2} \text{sgn}(\text{fric}_0)$$

So we write the value of the velocity $u_2$ at the interface as follows:

$$u_{2|z=\eta} = \frac{1}{\varepsilon^2}\, u_1^0 - \frac{1}{\varepsilon^2 \sqrt{C_1^0}} |\text{fric}_0|^{1/2} \text{sgn}(\text{fric}_0)$$

Thus, (87) turns into

$$\tilde{u}_2(z) = \frac{1}{\varepsilon^2}\, u_1^0 - \frac{1}{\varepsilon^2}\, \frac{\tilde{h}_m^{1/2}}{\sqrt{rC_1^0}Fr_1} |\mathcal{P}|^{1/2} \text{sgn}(\mathcal{P}) \tag{91}$$

with

$$\mathcal{P} = \Big( \nabla_x(rh_1 + h_2 + b) + (1-r)\text{sgn}(u_2)\tan\delta \Big).$$

Since the relation (82) is valid for any friction laws, the completion of the equation for the layer 1 does not change and is given by (89). The dimension form for the complete system is given by (17)-(20).

# B   Energy balance

## B.1   Proof of Theorem 2.1

In this appendix we prove the result of the energy balance associated to the proposed system that is held in Theorem 2.1. Concretely, we prove that the deduced model verifies an exact entropy dissipation energy.

For convenience, we remind here the proposed model given by equation (17)

$$\begin{cases} \partial_t h_1 + \text{div}_x q_1 = 0, \\[2mm] \partial_t q_1 + \text{div}_x(h_1(u_1 \otimes u_1)) + \dfrac{1}{2}g\nabla_x h_1^2 + gh_1\nabla_x(b + h_2) + \dfrac{gh_m}{r}\mathcal{P} = 0, \\[2mm] \partial_t h_2 + \text{div}_x\left( h_m\, v_b\, \sqrt{(1/r - 1)gd_s} \right) = 0, \\[2mm] \partial_t h_f = -T_m. \end{cases}$$

with
$$\mathcal{P} = \left( \nabla_x (rh_1 + h_2 + b) + (1-r)\mathrm{sgn}(u_2)\tan\delta \right)$$

and

- for the model $(LF)$,
$$v_b = v_b^{(LF)} = \frac{1}{\sqrt{(1/r-1)gd_s}} u_1 - \frac{\vartheta}{1-r}\mathcal{P}.$$

- for the model $(QF)$,
$$v_b = v_b^{(QF)} = \frac{1}{\sqrt{(1/r-1)gd_s}} u_1 - \sqrt{\frac{\vartheta}{1-r}}\, |\mathcal{P}|^{1/2}\mathrm{sgn}(\mathcal{P}).$$

In the following lines, we give the proof of the Theorem 2.1.

First we multiply the momentum equation for the layer 1 by $ru_1$ and we use the mass equation to obtain:

$$\frac{r}{2}\partial_t(gh_1^2 + h_1|u_1|^2) + r\mathrm{div}_x\left(h_1u_1\left(\frac{|u_1|^2}{2} + gh_1\right)\right) + \underbrace{rgh_1u_1\nabla_x(b+h_2)}_{(a)} + \underbrace{gh_mu_1\mathcal{P}}_{(b)} = 0$$

Now we multiply the equation for the layer 2 by $\tilde{\mathcal{P}} = g\big((rh_1+h_2+b) + x\,(1-r)\mathrm{sgn}(u_2)\tan\delta\big)$. Note that $\nabla_x\tilde{\mathcal{P}} = g\mathcal{P}$.

- For the linear friction law case,
$$\underbrace{\tilde{\mathcal{P}}\partial_t h_2}_{(c)} + \underbrace{\tilde{\mathcal{P}}\mathrm{div}_x(h_mu_1)}_{(d)} - \underbrace{\tilde{\mathcal{P}}\mathrm{div}_x\left(h_m\frac{\vartheta}{1-r}\mathcal{P}\,\sqrt{(1/r-1)gd_s}\right)}_{(e^L)} = 0;$$

- for the quadratic friction law case
$$\underbrace{\tilde{\mathcal{P}}\partial_t h_2}_{(c)} + \underbrace{\tilde{\mathcal{P}}\mathrm{div}_x(h_mu_1)}_{(d)} - \underbrace{\tilde{\mathcal{P}}\mathrm{div}_x\left(h_m\sqrt{\frac{\vartheta}{1-r}}|\mathcal{P}|^{1/2}\mathrm{sgn}(\mathcal{P})\,\sqrt{(1/r-1)gd_s}\right)}_{(e^Q)} = 0.$$

We use the definition of $\tilde{\mathcal{P}}$ to decompose the term $(c)$ in two parts, $(c) = (c)_1 + (c)_2$:

$$(c)_1 = g(rh_1 + h_2 + b)\partial_t h_2$$

$$(c)_2 = gx\,(1-r)\mathrm{sgn}(u_2)\tan\delta\,\partial_t h_2 = \partial_t\big(gh_2\,x\,(1-r)\mathrm{sgn}(u_2)\tan\delta\big)$$

Now, after some simple calculations, the terms $(a)$ and $(c)_1$ gives:

$$(a) + (c)_1 = \frac{1}{2}g\partial_t\big((b+h_2)^2\big) + \text{div}_x(rgh_1(b+h_2)u_1) + rg\partial_t\big(h_1(b+h_2)\big)$$

Finally

$$(b) + (d) = \text{div}_x(h_m u_1 \tilde{\mathcal{P}})$$

and the last terms read

$$(e^L) = -\text{div}_x\left(h_m\frac{\vartheta}{1-r}\tilde{\mathcal{P}}\mathcal{P}\sqrt{(1/r-1)gd_s}\right) + gh_m\frac{\vartheta}{1-r}\sqrt{(1/r-1)gd_s}\,|\mathcal{P}|^2$$

or

$$(e^Q) = -\text{div}_x\left(h_m\sqrt{\frac{\vartheta}{1-r}}\tilde{\mathcal{P}}|\mathcal{P}|^{1/2}\text{sgn}(\mathcal{P})\sqrt{(1/r-1)gd_s}\right) + gh_m\sqrt{\frac{\vartheta}{1-r}}\sqrt{(1/r-1)gd_s}\,|\mathcal{P}|^{3/2}$$

So, finally we obtain (43) and (44), where the right hand side in both cases are non-positive.

## B.2  Proof of Theorem 2.2

In this section we analyze the energy associated to a general Saint-Venant-Exner model for a modified friction in the momentum equation of the Saint-Venant system:

$$\tau/\rho_1 = g\varphi(h_1)u_1|u_1| + \frac{1}{r}\xi_m\sqrt{g\varphi(h_1)}\,\mathcal{R}.$$

This model is given in equations (1)-(3) that we remind next:

$$\begin{cases} \partial_t h_1 + \text{div}_x q_1 = 0, \\[2mm] \partial_t q_1 + \text{div}_x\left(\frac{|q_1|^2}{h_1} + \frac{1}{2}gh_1^2\right) + gh_1\text{div}_x(h_2+b) + \tau/\rho_1 = 0, \\[2mm] \partial_t h_2 + \text{div}_x q_b = 0, \end{cases}$$

where

$$\frac{q_b}{Q} = \frac{1}{1-\varphi}\text{sgn}(\tau)\,k_1\,\theta^{m_1}\,(\theta - k_2\theta_c)_+^{m_2}\,(\sqrt{\theta} - k_3\sqrt{\theta_c})_+^{m_3},$$

$$Q = d_s\sqrt{g(1/r-1)d_s}, \quad \theta = \frac{|\tau|d_s^2}{g(\rho_2-\rho_1)d_s^3} \quad \text{and} \quad \tau = \rho_1 g\varphi(h_1)u_1|u_1|.$$

To prove Theorem 2.2 we follow the same development above. First, we write the discharge $q_b$ in the following way:

$$\frac{q_b}{Q} = \xi_m\text{sgn}(\tau)\Big(\sqrt{\theta} - \sqrt{\theta_c} + \text{sgn}(\tau)g\nabla_x(rh_1+h_2+b)\Big),$$

where

$$\xi_m = \frac{1}{1-\varphi}\, k_1\, \theta^{m_1}\, (\theta - k_2\theta_c)_+^{m_2}\, (\sqrt{\theta} - k_3\sqrt{\theta_c})_+^{m_3}\, \frac{1}{(\sqrt{\theta} - \sqrt{\theta_c} + \mathrm{sgn}(\tau)g\nabla_x(rh_1 + h_2 + b))}.$$

Now, from the definition of $\theta$ and $\tau$ we write:

$$\theta = \frac{\rho_1\varphi(h_1)|u_1|^2}{(\rho_2 - \rho_1)d_s}, \quad \text{so} \quad \sqrt{\theta} = \frac{\sqrt{\varphi(h_1)}|u_1|}{\sqrt{(1/r - 1)d_s}}$$

with

$$\mathcal{R} = \mathrm{sgn}(\tau)\sqrt{\theta_c} - g\nabla_x(rh_1 + h_2 + b).$$

So, finally

$$q_b = \xi_m v_b \sqrt{g(1/r - 1)d_s} \quad \text{with} \quad v_b = \frac{\sqrt{g\varphi(h_1)}}{\sqrt{g(1/r - 1)d_s}}u_1 - \mathcal{R}.$$

And we can write the equation for $h_2$ as follows:

$$\partial_t h_2 + \mathrm{div}_x\Big(\xi_m\sqrt{g\varphi(h_1)}u_1 - \xi_m\mathcal{R}\sqrt{g(1/r - 1)d_s}\Big) = 0.$$

Note that we have rewritten the evolution equation for the sediment in a different manner but no modification has been introduced in it. As we mentioned above, the only modification needed to obtain the dissipative energy balance is taken into account in the friction term in the momentum conservation equation of the Saint-Venant system.

First, we multiply the momentum equation of the first layer by $ru_1$ and we use the mass equation to obtain:

$$\frac{r}{2}\partial_t\big(gh_1^2 + h_1|u_1|^2\big) \;+\; r\mathrm{div}_x\left(h_1u_1\left(\frac{|u_1|^2}{2} + gh_1\right)\right) + \underbrace{rgh_1u_1\nabla_x(b + h_2)}_{(a)}$$

$$+\;\; g\varphi(h_1)|u_1|^3 + \underbrace{\xi_m\sqrt{g\varphi(h_1)}\mathcal{R}u_1}_{(b)} = 0$$

Now we multiply the equation for layer 2 by $\tilde{\mathcal{R}} = \mathrm{sgn}(\tau)\big(g(rh_1 + h_2 + b) + x\sqrt{\theta_c}\big)$, where $\nabla_x\tilde{\mathcal{R}} = \mathcal{R}$. Then,

$$\underbrace{\tilde{\mathcal{R}}\partial_t h_2}_{(c)} + \underbrace{\tilde{\mathcal{R}}\mathrm{div}_x\Big(\xi_m\sqrt{g\varphi(h_1)}u_1\Big)}_{(d)} - \underbrace{\tilde{\mathcal{R}}\mathrm{div}_x\Big(\xi_m\mathcal{R}\sqrt{(1/r - 1)gd_s}\Big)}_{(e)} = 0.$$

In the same manner than before, we obtain that

$$(a) + (c) = \frac{1}{2}g\partial_t\big((b + h_2)^2\big) + \mathrm{div}_x(rgh_1(b + h_2)u_1) + rg\partial_t\big(h_1(b + h_2)\big) - \partial_t\big(h_2\mathrm{sgn}(\tau)\sqrt{\theta_c}\big)$$

Finally

$$(b) + (d) = \text{div}_x(\xi_m u_1 \sqrt{g\varphi(h_1)} \tilde{\mathcal{R}})$$

and the last term reads

$$(e) = -\text{div}_x \left( \xi_m \tilde{\mathcal{R}} \mathcal{R} \sqrt{(1/r - 1)g d_s} \right) + \xi_m \sqrt{(1/r - 1)g d_s} |\mathcal{R}|^2$$

Then, (45) is satisfied.